\documentclass[a4paper,11pt]{article}
\pdfoutput=1
\usepackage{jcappub} 
\usepackage{footnote}	
\usepackage{xcolor}
\usepackage{bigints}
\usepackage{multirow}
\usepackage{appendix}
\usepackage[normalem]{ulem}
\usepackage{bm,relsize}        
\usepackage{amsfonts,amsmath,amssymb,amsthm,mathtools}
\def\DM		{{\mathsmaller{\text{DM}}}}

\graphicspath{{figs/}} 

\begin{document}

\title{511 keV Galactic Photons from a Dark Matter Spike }

\author[a,b]{Pedro De la Torre Luque,}
\author[c]{Shyam Balaji,}
\author[c]{Malcolm Fairbairn}
\author[d,1]{Filippo Sala\note{On leave of absence from LPTHE, CNRS \& Sorbonne Universit\'{e}, Paris, France.}}
\author[e,f,g]{and Joseph Silk}

\affiliation[a]{Departamento de F\'{i}sica Te\'{o}rica, M-15, Universidad Aut\'{o}noma de Madrid, E-28049 Madrid, Spain}
\affiliation[b]{Instituto de F\'{i}sica Te\'{o}rica UAM-CSIC, Universidad Aut\'{o}noma de Madrid, C/ Nicol\'{a}s Cabrera, 13-15, 28049 Madrid, Spain}
 \affiliation[c]{Physics Department, King’s College London, Strand, London, WC2R 2LS, United Kingdom}
 \affiliation[d]{Dipartimento di Fisica e Astronomia, Università di Bologna
and INFN sezione di Bologna, Via Irnerio 46, I-40126 Bologna, Italy}
\affiliation[e]{Institut d’Astrophysique de Paris, UMR 7095 CNRS \& Sorbonne Universit\'{e}, 98 bis boulevard Arago, F-75014 Paris, France}
\affiliation[f]{Department of Physics and Astronomy, The Johns Hopkins University, 3400 N. Charles	Street, Baltimore, MD 21218, U.S.A.}
\affiliation[g]{Beecroft Institute for Particle Astrophysics and Cosmology, University of Oxford, Keble	Road, Oxford OX1 3RH, U.K.}

\emailAdd{pedro.delatorre@uam.es}
\emailAdd{shyam.balaji@kcl.ac.uk}
\emailAdd{malcolm.fairbairn@kcl.ac.uk}
\emailAdd{f.sala@unibo.it}
\emailAdd{silk@iap.fr}

\abstract{
We propose that a dark matter (DM) spike around the Galactic Center's (GC) supermassive black hole, Sgr A*, could account for most of the bulge's measured 511 keV line intensity while remaining cosmologically compatible. DM annihilation can be the primary source of the 511 keV line emission without violating constraints from disk emission observations and in-flight positron annihilation with the interstellar medium, provided the disk emission is dominated by an astrophysical source of low-energy positrons. We find that a DM mass up to approximately 20 MeV, either with a Gondolo-Silk spike or one softened by stellar heating, could explain the observed 511 keV bulge emission profile. 
Our proposal can be tested by future observations of the continuum diffuse emission close to the GC.
}

\maketitle

\section{Introduction}
Early observations of the diffuse $\gamma$-ray emission at MeV scales~\cite{1972ApJ...172L...1J, 1973ApJ...184..103J, 1986ApJ...302..459L} revealed the presence of a bright line at $511$~keV, from $e^+e^-$ annihilations via para-positronium, that could not be explained solely by the positrons created by cosmic-ray (CR) collisions with the interstellar medium (ISM). In fact, these observations pointed to a continuous positron injection from sources in the Galactic bulge. Recent measurements have detected the $511$~keV bulge emission with a significance exceeding $50 \sigma$ and have 
identified a much more extended emission coinciding with the Galactic disk~\cite{Siegert:2015knp}.
In addition, spectral analyses of the $511$~keV line have found high significance in the detection of continuum emission associated with ortho-positronium (the triplet state of positronium~\cite{KARSHENBOIM_2004, Badertscher_2007}) and even in-flight positron annihilation (direct $e^+$-$e^{-}$ annihilation of high energy positrons), revealing that positron sources contribute significantly to the diffuse $\gamma$-ray flux at $\sim$MeV energies~\cite{Beacom:2005qv, sizun2007constraints}. 


Different sources have been proposed to explain the emission from the disk. Sources explaining contributions to the disk emission can be CR interactions, pulsars injecting $e^{\pm}$, or sources synthesizing radioactive elements (like $^{26}$Al in massive stars, $^{44}$Ti in core-collapse supernovae or $^{56}$Ni in supernovae 1A) that decay into positrons ($\beta+$ decays)~\cite{Siegert:2021trw, Frontera}. Massive stars have received special attention since their emission can be observationally inferred and could be able to explain a significant portion of the disk emission~\cite{Prantzos_2011, Knoedlseder_2005, Weidenspointner_2006}. However, the measured bulge emission requires a spatial morphology and an injection rate that does not seem to easily fit with known candidates, such as low-mass X-ray binaries, type 1A supernovae or other sources expected to be located around the Galactic Center (GC)~\cite{Knoedlseder_2005, Weidenspointner_2006, Frontera, Prantzos_2011, Rev511}. 

Ref.~\cite{Boehm:2003bt} suggested that sub-GeV dark matter (DM) could be the dominant source of the bulge emission. Then, Ref.~\cite{Ascasibar_2006} showed that DM decay cannot be responsible for the bulge emission, whilst velocity-independent DM annihilation could be consistent with the $511$~keV bulge emission for DM density distributions following a Navarro-Frenk-White (NFW) distribution~\cite{Navarro:1995iw}.
However, constraints from in-flight positron annihilation imply that, for a positron source (of a few degrees of angular size) to explain the 511~keV bulge emission, positrons need to be injected with an energy of at most $\sim 3$~MeV~\cite{Beacom:2005qv}.
This in turn implies that the mass of DM annihilating into $e^+e^-$ must satisfy $M_\DM \lesssim 3$~MeV, unless one invokes cascade annihilations~\cite{Jia:2017iyc} or heavier `exciting' DM with huge cross sections~\cite{Finkbeiner:2007kk,Cappiello:2023qwl}. Such small masses were found to be in conflict with cosmological data, so that an MeV DM explanation for the 511~keV bulge emission had been claimed excluded in 2016~\cite{Wilkinson:2016gsy}.
Later the analysis of these cosmological constraints have been refined, in particular Refs.~\cite{Escudero:2018mvt,Sabti:2019mhn} showed that DM masses down to $\sim 1$~MeV can be made consistent with CMB and Big Bang Nucleosynthesis (BBN) observations, as long as there is a small additional neutrino component injected at early times concurrent with the electron-positron contribution from DM. Ref.~\cite{Ema:2020fit} then connected this observation with the 511~keV line and built models of DM, annihilating into both $e^+e^-$ and neutrinos, that explain the 511~keV bulge emission without being excluded by any other known laboratory, astrophysical or cosmological constraint.

Moreover, the calculations of the $511$~keV emission mentioned above did not take into account that positrons injected with energies above a few MeV can propagate appreciably from the injection source, thus smearing the expected $511$~keV profile with respect to the DM density profile~\cite{DelaTorreLuque:2023cef}. In fact, Ref.~\cite{DelaTorreLuque:2023cef} showed that a NFW profile is not able to explain the $511$~keV longitude profile when including propagation of the positrons with state-of-the-art propagation parameters.
Ref.~\cite{DelaTorreLuque:2023cef} also pointed out that the observed disk emission (and, in particular, at high longitudes) strongly constrains the DM production of $e^\pm$, making it very difficult to explain the observed bulge emission (i.e. at the central longitudes) without exceeding the observations of the disk at high longitudes, and that when including the effects of propagation, in a similar way as we do in this work, a slope around $\gamma\gtrsim1.1$ would be necessary to explain the full $511$~keV line profile (however, if no broadening of the signal by diffusion is considered, a profile with $\gamma=1$ notably reproduces the morphology of the line). 
This challenges the scenario where DM is the main culprit for bulge emission of the $511$~keV line: not only one needs to build models with an ad-hoc injection of neutrinos at early times, but one needs DM density profiles steeper than what suggested by numerical simulations. 

In this paper, we demonstrate that one can explain most of the observed intensity of the $511$~keV line at the bulge in terms of DM annihilations into $e^+e^-$, and without the need of an ad-hoc injection of neutrinos in the early universe, if Sgr A* has had sufficient time to accrete DM in the form of a density `spike' that could subsequently produce an enhanced annihilation signal.
In particular, we show that both the observed latitudinal and longitudinal profiles can be reasonably explained for a $10$-$20$ MeV DM for DM distributions forming a spike around Sgr A*, as long as the disk emission is being generated by one or more non-DM sources, something that is expected.
We demonstrate that this is possible not only for the `Gondolo-Silk' DM spikes~\cite{Gondolo:1999ef} where DM accretes adabatically on SgrA* unperturbed, but also for the more conservative spikes that are softened because of DM heating by stars in the vicinity of Sgr A*, see e.g.~\cite{Balaji:2023hmy}.

This paper is organized as follows: In Sec.~\ref{sec:method}, we cover the main arguments motivating the use of a DM distribution with a spike around the central black hole, then we explain how we simulate the positron injection and propagation from annihilation of sub-GeV DM along with the calculation of the $511$~keV line and the associated continuum emission, detailed in the same section.
We show comparisons of our predictions with current MeV data in Section~\ref{sec:results} and discuss their implications in Sec.~\ref{sec:Conclusion}.

\section{511 keV and continuum signals from DM with a central spike}
\label{sec:method}

\subsection{DM distributions with a spike around Sgr A*}
\label{sec:DM}
The Sgr A* gravitational potential may significantly enhance the DM density in its vicinity, in contrast to the outer halo region. This increased DM concentration near Sgr A* is termed a `spike'. The initial suggestion to observe this spike through telescopes to detect stronger DM annihilation signals was made decades ago~\cite{Gondolo:1999ef}. Since then, our understanding of physics of the spike has greatly advanced.
In this section, we utilize this improved knowledge to establish and justify benchmarks for the DM spike profile, which will serve as an indication of astrophysical uncertainties on the DM annihilation signal intensity.

For the Milky Way's DM mass density as a function of $r$, the distance from Sgr A*, we adopt the following parametrization~\cite{Balaji:2022wqn}
\begin{align}
\label{eq:spikedensity}
    \rho(r)= \Big(1- \frac{2 R_S}{r}\Big)^{3/2} \times \left\{
        \begin{array}{ll}
            0 & \quad r< 2  R_S \\
    \displaystyle\rho_\textrm{sat}\Big(\frac{r}{R_{\rm sat}}\Big)^{-0.5} & \quad 4 R_S \leq r < R_\textrm{sat} \\
            \displaystyle\rho_\textrm{spike}(r) & \quad R_\textrm{sat} \leq r < R_\textrm{sp} \\
            \rho_\textrm{halo}(r) & \quad r \geq R_\textrm{sp}
        \end{array} ,
    \right.
\end{align}
where $R_S=2GM_\textrm{BH}/c^2=2.95\,(M_\textrm{BH}/M_\odot)\textrm{\,km}$ is the Schwarzschild radius of the supermassive black hole (SMBH), $R_\textrm{sp}$ and $\rho_\textrm{spike}$ are the radial extension and the mass density profile of the DM spike respectively,  $R_\textrm{sat}$ and $\rho_\textrm{sat}$ are the saturation radius and density of the spike due to DM annihilation respectively.
$R_\textrm{sat}$ is defined by $\rho_\textrm{sat}=\displaystyle\rho_\textrm{spike}(R_\textrm{sat})$, where
    \begin{align}
    \label{eq:rhosat}
    \rho_{\rm sat} & = \frac{m_\chi}{\langle\sigma v\rangle t_{\rm BH}} \nonumber \\
    &\simeq 3.17\times\,10^{11}\,{\rm GeV}\,{\rm cm}^{-3}\frac{m_\chi}{10~{\rm TeV}} \frac{10^{-25} \textrm{cm}^3/\text{s}}{\langle\sigma v\rangle}\frac{10^{10}\textrm{yr}}{t_{\rm BH}}\,,
\end{align}
where we will use $t_\textrm{BH}=10^{10}$yr as the age of Sgr A*. At $r<R_{\rm sat}$ the density continues to grow, rather than forming a plateau, due to DM particles'orbits being non-circular. The slope $\sim r^{-0.5}$ below the saturation radius is found under the assumption of $s$-wave DM annihilations~\cite{Vasiliev:2007vh,Shapiro:2016ypb}.
We model the spike density profile as
\begin{equation}
\label{eq:Rsat}
\rho_\textrm{spike}(r)
=\displaystyle\rho_\textrm{halo}(R_\textrm{sp})\left(\frac{r}{R_\textrm{sp}}\right)^{-\gamma_\textrm{sp}(r)}\,,
\end{equation}
where $\gamma_\textrm{sp}(r)$ is the slope of the spike, for which we will make different benchmark choices.
The halo distribution is defined with the usual NFW~\cite{Navarro:1995iw} profile, given as
\begin{equation}
\label{eq:NFW}
    \rho_\textrm{halo}^\textrm{NFW}(r)
    = \rho_s\left(\frac{r}{r_s}\right)^{-1} \left(1+\frac{r}{r_s}\right)^{-2} \,,
\end{equation}
with $r_s=18.6\,$kpc and $\rho_s\,=\,\displaystyle\rho_\odot(R_\odot/r_s) (1+R_\odot/r_s)^{2}$, where $R_\odot\,=\,8.2\,\,{\rm kpc}$ is the sun position and $\rho_\odot=0.42\, \textrm{GeV}/\textrm{cm}^3$ is the local DM density~\cite{Pato_2015}.
Finally, the prefactor $(1- 2 R_S/r)^{3/2}$ accounts for DM capture by the BH~\cite{Gondolo:1999ef}, corrected in light of the relativistic treatment of~\cite{Sadeghian:2013laa} (the inclusion of this prefactor is the only improvement here with respect to the profile's parametrization of~\cite{Balaji:2022wqn}). In this paper we do not consider halo profiles more flat than NFW because, to start with, we aim at explaining a peaked signal, nor we consider the possibility that DM self-interact to an extent that it also affects its density~\cite{Rocha:2012jg,Kaplinghat:2013xca}.

Refs.~\cite{Gnedin:2003rj,Merritt:2003qc} showed that the spike begins to grow around $R_{\rm sp} \simeq 0.2\,R_h$, where $R_h = GM_{\rm BH}/v_o^2$ is the radius of gravitational influence of the SMBH and $v_o$ is the dispersion in velocity of the stars populating the inner halo.
Using $M_{\rm BH}\,=\,4.3\times 10^6\,{\rm M}_\odot$ for Sgr A$^*$~\cite{GRAVITY:2021xju} and $v_o\,=\, 105 \pm 20\, {\rm km\, s}^{-1}$~\cite{2009ApJ...698..198G}, we obtain $R_h\,=\,1.7\,{\rm pc}$ and
    \begin{equation}
    \label{eq:Rspike}
        R_{\rm sp}\,\simeq\,0.34\,{\rm pc}\,.
    \end{equation}

We consider three benchmark spike scenarios, which we encode in the slope $\gamma_\textrm{sp}(r)$
\begin{itemize}
    \item[$\diamond$] \textbf{Gondolo-Silk (GS)}. The spike slope is given by~\cite{Gondolo:1999ef,Ullio:2001fb} 
    \begin{equation}
    \gamma_{\rm sp}^{\rm GS} = \frac{7}{3}\,,
    \end{equation}
    it follows from the assumptions of i) adiabatic accretion of a DM halo around a BH at its (peaked) center ii) without any significant perturbation from other effects, like gravitational interactions with stars possibly surrounding the BH.\footnote{\label{foot:Yu}
    Note that the use of $v_o\,=\, 105 \pm 20\, {\rm km\, s}^{-1}$ in our determination of $R_h$ is valid only if the dispersion in DM velocities is the same as in stars, and for the GS profile this is in principle not valid as the two populations evolve independently. This would lead to a larger $R_{\rm sp}$ for the GS profile. We still use Eq.~(\ref{eq:Rspike}) both for being conservative and to respect the GRAVITY constraints discussed later in this section.}
    \item[$\diamond$] \textbf{Maximal stellar heating (*MAX)}.
    The stellar heating benchmark is a modification of the GS one caused by the presence of baryonic matter, such as stars, in the vicinity of Sgr A*. The gravitational interplay between DM and these baryons tends to mitigate the spike density. For example, the gravitational `heating' caused by the nuclear star cluster—that is, the stars located within the central parsecs of the Milky Way—could notably reduce the DM spike. This interaction leads to a balanced spike configuration that can be as low as~\cite{Gnedin:2003rj,Bertone:2005hw,Merritt:2006mt,Shapiro:2022prq} 
    \begin{equation}
        \gamma_{\rm sp}^{\rm *MAX} = 1.5\,,
    \end{equation}
    basically following the measured~\cite{2020Gallego} slope of the nuclear star cluster's density.
    This model is denoted as *MAX because it represents the case with ``maximal" stellar heating. Indeed, it is not unreasonable to conceive that gravitational heating of the DM spike should be stop being efficient at the small radii where there are no stars.
    \item[$\diamond$] \textbf{Minimal stellar heating (*MIN)}.
    We therefore also consider another scenario for DM spike softening, where
    \begin{align}
    \label{eq:gamma_starMIN}
    \gamma_{\rm sp}^{\rm *MIN}
    = \left\{
        \begin{array}{ll}
            \dfrac{7}{3} & \quad r < 0.01~{\rm pc} \\
            1.5 & \quad r \geq 0.01~{\rm pc}
        \end{array} ,
    \right.
\end{align}
    Based on stellar density data from the inner 0.04 pc × 0.04 pc region~\cite{2019Habibi,2020Gallego}, the mean separation between nuclear cluster stars is about 0.01 parsecs, suggesting minimal scattering between DM and stars within this region, thus preserving the DM spike. This scenario, which we denote *MIN, corresponds to the minimal stellar heating scenario in this work.\footnote{Note that the nomenclature in this work is different from our previous work; our *MAX and *MIN profiles here correspond to the profiles denoted $\bigstar$ heating and $\bigstar$ heating- respectively in Ref.~\cite{Balaji:2023hmy}.} This assumes the black hole grew mainly before the nuclear star cluster formed, decoupling the DM and stellar profiles. The possible origin of nuclear star clusters through globular cluster mergers further supports this decoupling,
preventing the softening of the DM spike due to the reduced numbers of older/brighter giants near the GC \cite{2019Habibi,2020Gallego}.
A spike slope as in Eq.~(\ref{eq:gamma_starMIN}) is also found in simulations of the evolution of the DM spike that indeed account for gravitational heating~\cite{Bertone:2005hw}, corroborating our choice of this benchmark.
\end{itemize}
The study of stellar orbits around Sgr A*, using the GRAVITY instrument since 2016~\cite{GRAVITY:2024tth}, has provided unprecedented precision in tracking these orbits. By analyzing astrometric and spectroscopic data from stars like S2, S29, S38, and S55, the collaboration has recently constrained any extended mass distribution around Sgr A* to be at most  $1200 M_\odot$ up to the S2 apocenter (which is around $10^{-2}$ pc from the GC).
This extended mass includes astrophysical objects (like white dwarfs and neutron stars) that, according to simulations, can contribute to it at a level close to the GRAVITY upper limit. 
The extended mass includes also a possible DM spike, thus potentially constraining our benchmarks.
The cuspiest case considered in this work, i.e. the Gondolo-Silk spike model, has enclosed mass $\lesssim 40 M_\odot$ which easily complies with this constraint. 
The DM mass enclosed within $10^{-2}$ pc from the GC, for a GS spike, would surpass a few hundreds $M_\odot$ for $R_{\rm spike} >$~few~pc (in particular, for our choice of DM density profiles, the enclosed mass within this region is 0.129287, 26.941, 1.9746, 1.00483 M$_{\odot}$ for the an NFW, Gondolo-Silk, *Min and *Max, respectively), thus challenging the possibility of a much larger $R_{\rm sp}$ for the GS case (see footnote~\ref{foot:Yu}).

\begin{figure}[t]
\centering
\includegraphics[width=0.49\linewidth]{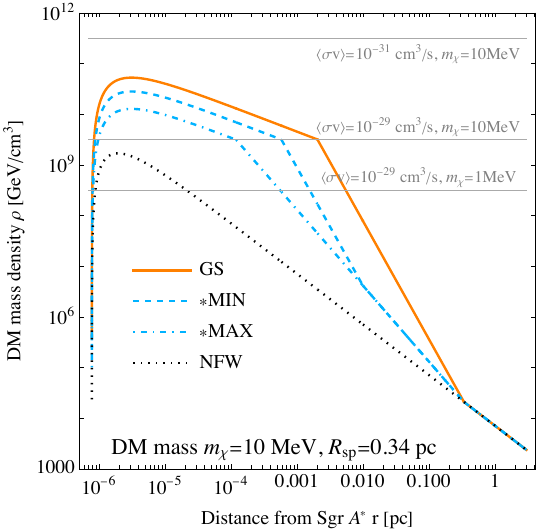}
\includegraphics[width=0.49\linewidth]{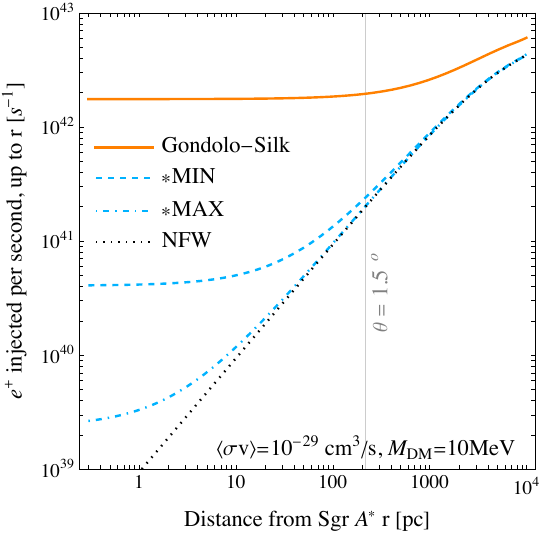}
\caption{The DM density as a function of radial positions from the GC is shown on the left, as modelled by Eq.~(\ref{eq:spikedensity}). The rate of positrons injected by DM in a sphere of radius $r$ is shown on the right, as obtained from the volume integral of Eq.~(\ref{eq:source}).
In both cases we display the following density profiles: NFW (black dotted), Gondolo-Silk spike (GS, orange), GS softened by star-heating down to the inner stars (*MIN, dashed blue) and down to Sgr A* (*MAX, dash-dotted blue). We take a DM mass of $m_\chi=10$ MeV and an annihilation cross section of $\langle \sigma v\rangle = 10^{-29}$~cm$^3$/s. For reference, in the left-hand plot we also display the saturation of DM density induced by different benchmark values of $m_\chi$ and $\sigma v$.}
\label{fig:density_injection}
\end{figure}

 In Fig.~\ref{fig:density_injection}, we display in the left-hand panel the radial density profiles for the GS, *MIN and *MAX spikes, and the usual NFW for comparison.
 In the right-hand panel we show the total number of injected positrons up to a certain radius, for the same DM profiles. One reads there that, for any given spike, the injection up to a certain radius is dominated by the one at the spike rather than by the diffuse NFW one. These injected positrons must then be propagated, as we will discuss in the next section.
We will show that DM spike profiles are expected to produce a negligible amount of disk emission, while being bright enough at the central longitudes to explain most of the $511$~keV bulge emission. 
Finally, we also note that the rotation of Sgr A* (see e.g.~\cite{Daly:2023axh}) would not affect the formation of the DM spike~\cite{Ferrer:2017xwm}.

\subsection{Positron injection and propagation}
\label{sec:PosGal}

Given that our goal is to evaluate the diffuse $511$~keV emission as precisely as possible, we simulate the evolution of the position and energy of positrons with a state-of-the-art propagation set-up and approximate the thermalized distribution of positrons as the steady-state diffuse positron distribution generated from DM annihilations. In this way, we compute the $511$~keV line emission as proportional to the thermalized distribution of positrons. 
Following a similar procedure as in Ref.~\cite{DelaTorreLuque:2023olp}, we use a customized version~\cite{de_la_torre_luque_2023_10076728} of the {\tt DRAGON2} code~\cite{DRAGON2-1, DRAGON2-2}, a dedicated CR propagation code prepared to simulate CR diffusion, accounting for all diffusion-reacceleration-advection-loss effects in the propagation of galactic CRs~\cite{Ginz&Syr}. We use the same expression for the diffusion coefficient and propagation parameters (the best-fit values obtained from combined fits to AMS-02 CR data) as in Ref.~\cite{DelaTorreLuque:2023olp}, to which we refer the reader for more details. 
The injection per unit volume of $e^{\pm}$ particles from annihilating self-conjugate DM is given by the source term
%
\begin{equation}
     Q_e (\vec{x},E_e) =
\frac{\langle \sigma v\rangle}{2} \left(\frac{\rho_\chi(\vec{x})}{m_\chi}\right)^2\frac{dN_e^{\textrm{ann}}}{dE_e} \,,
\label{eq:source}
\end{equation}
where $\rho_\chi(\vec{x})$ is the DM energy density at the position $\vec{x}$ and $\frac{dN_e^{\textrm{ann}}}{dE_e} = \delta(E-m_{\chi})$ is the positron injection yield in the direct $\chi \chi \rightarrow e^+ e^-$ annihilation channel. We set the normalization of the DM density imposing that its density at Earth position is $\rho_\chi(\vec{x}=\vec{x_{\odot}})=0.42$~GeV/cm$^3$. 
With this source term we solve the propagation equation  of electrons and positrons numerically, as done, e.g., in Ref.~\cite{DelaTorreLuque:2023olp}. In this way, we calculate the steady-state e$^{\pm}$ distribution and energy spectrum in the Galaxy.

Since the spike models that we use are characterized by an abrupt increase of DM density towards the GC, we have updated the code to be able to deal with a logarithmic spatial grid.
Unfortunately, our propagation code cannot reach resolutions below $\sim 0.01$~pc, because of numerical stability. Therefore, we adopt a spatial grid with a bin size varying from $0.01$~pc in the inner Galaxy to $\simeq 200$~pc, increasing progressively. 
This resolution easily suffices to study the spike models presented in Ref.~\cite{Balaji:2023hmy}. We model the inner spatial bin by integrating the DM density upto the radial edge of this bin and evaluating the subsequent $e^{\pm}$ emission from this bin. 
We simulate electron-positron signals from DM annihilation in the range of kinetic energies from $100$~eV to $1$~GeV, with an energy resolution of $5\%$.
Additionally, we remark that in these simulations  account for all the sources of energy losses, inelastic interactions, triplet pair production~\cite{Gaggero:2013eik} and in-flight annihilation of positrons.

The propagation of positrons at sub-MeV energies remains poorly understood, particularly in the sub-keV range. At these low energies, the standard diffusion model may break down, potentially preventing positrons from propagating over significant distances. Additionally, if positrons experience nearly catastrophic energy losses or if diffusion mechanisms operate in an unconventional manner, their propagation could be severely limited, possibly confining them to scales below a parsec. These uncertainties make it challenging to reliably model positron behavior at sub-MeV energies, which is why such cases are often excluded from consideration. However, for the sake of having a reasonable estimate, we use an extrapolation of the diffusion coefficient that is set from CR observations at higher energies.
Under this assumption, positrons with energies above the MeV-scale can propagate distances sufficiently far, before losing their energy and thermalizing, that their associated $511$~keV emission must be seen at a few degrees away from the GC. Especially, in the case where reacceleration is taken into account, $\sim1$~MeV positrons could propagate hundreds of parsecs (see Fig.~\ref{fig:MeanFreePath} in Appendix~\ref{sec:App_Uncerts}). However, the conditions in the inner Galaxy are quite complex, in comparison to other Galactic regions, and the diffusion of charged particles may be significantly different. As an example, Ref.~\cite{Pos_transport} found that $\sim1$~MeV positrons could propagate kpc distances before thermalizing, because positrons would have negligible interactions with magnetohydrodynamic waves. We discuss the impact of the choice of different reacceleration benchmarks in App.~\ref{sec:App_Uncerts}, and we anticipate that they do not affect the message of our paper.

\subsection{Line emission}
\label{sec:line}
From the distribution of diffuse positrons in the Galaxy as a function of position, $\phi^{e^+}(x,y,z)$, we can calculate the emission of $511$~keV $\gamma$-rays, from the decay of para-positronium, in every point of the Galaxy. A complete evaluation should consider that the $511$~keV production is also proportional to the total electron density $n_e$ (either from free electrons and electrons bound to the different atomic and molecular gas species and to dust grains), and the cross section of positron annihilation $\sigma^\textrm{ann}$ (through positronium formation from charge-exchange with hydrogen gas, which is expected to be the dominant mechanism of positronium production) at the energy of the thermalized positrons $E_\textrm{th}$ (see, e.g. Ref.~\cite{1991ApJ...378..170G} for more details). Positrons are produced from DM annihilations with $E = m_\chi \gg E_\textrm{th}$, but they thermalize with the medium before substantially annihilating with electrons and hence produce photons, see e.g. Refs.~\cite{Beacom:2005qv,DelaTorreLuque:2024zsr}. With these considerations, the $511$~keV line flux integrated over the line of sight takes the following form
\begin{equation}
    \frac{d\phi_{\gamma}^{511}}{d\Omega}=2k_{ps} \int ds \,\frac{d\phi^{e^+}}{d\Omega} \cdot n_e \cdot \sigma^\textrm{ann}(E_\textrm{th})\,,
    \label{eq:511line}
\end{equation}
where $k_{ps}=1/4$ is the fraction of positronium decays contributing to the $511$~keV line signal, $\phi^{e^+}$ is the steady-state flux of diffuse positrons at the thermal energy and the factor $2$ accounts for the emission of two $511$~keV photons per positron annihilation. 
$d\Omega=dl db \cos b$ is the solid angle element being $l$, $b$ and $s$ the galactic longitude, latitude and distance $s$ from the Earth contributing to the signal. 
We note that an additional component the direct annihilation of positrons with electrons, which also contributes to the total line emission. However, the fraction of positrons annihilating before reaching positronium state is found to be very small ($<5\%$~\cite{Jean_2005} - see more details in Sect.~\ref{sec:cont} too).

Given the uncertainties in the determination of the total electron density distribution (especially towards the GC) and since it is expected that it follows a smooth spatial distribution in the galactic disk without very high variations in the inner $\sim10$-$15$~kpc (roughly corresponding to the longitude range covered by the SPI data that we use here), we adopt the usual approach and estimate the distribution of the $511$~keV line emission as directly proportional to the spatial distribution of the diffuse positrons, thus setting the electron density to be uniform and with a value representative of the average electron density in the disk,  $n_e = 1$~cm$^{-3}$~\cite{1991Natur.354..121C}.
However, we expect that the electron density drops very quickly when moving out from the galactic plane, as the $511$~keV emission should as well. 
Observations by SPI have shown that the morphology of the 511 keV line favours the formation of positronium in an even mixture of warm hydrogen gas and electron gas. To capture this, we apply a scaling relation to the $511$~keV profiles, following the vertical distribution of an equal sum of warm gas and electron density in the Galaxy. In particular, for our benchmark evaluation, we implement a 50\% Nakanishi~\cite{Nakanishi_2003} (representing moderately warm gas medium) plus a 50\% Ferriere~\cite{Ferriere_1998} (for the electron density) to be consistent with these observations. Meanwhile, we adopt a constant electron density in the Galactic plane, as discussed above. In this way, our predicted $511$~keV longitude profile follows the distribution of thermalised positrons in the Galactic disk and the latitude profile follows the convolution of the vertical electron density distribution and the thermal positrons. We show the effects of convolving with the different radial gas distributions in the right panel of Fig.~\ref{fig:GS_Profs}, and discuss it in more detail in App.~\ref{sec:App_Uncerts}.  
 
Then, we approximate the total  $e^+ e^-$ annihilation cross sections, $\sigma^\textrm{ann}$, as the charge-exchange cross section with hydrogen~\cite{JeanP_2009}, that is about a million times greater than the direct annihilation of positrons with free electrons~\cite{Dirac}, and assume that positrons annihilate once they reach the thermal energy of a warm medium (with $T=8000$~K), as obtained in the analysis of the width of the $511$~keV line in Ref.~\cite{Jean_2005}.
Note that this implies that $n_e$ of Eq.~(\ref{eq:511line}) is dominated by the electrons in hydrogen for the purpose of computing the 511~keV signal.
Taking a different thermal energy would lead to change in the normalization of the signal, meaning our conclusions (except those on the values of best-fit DM annihilation cross sections) would remain unchanged.

On top of the $511$~keV DM emission, we add a disk component that reproduces emission from $\beta^+$ emitters in the disk (as massive stars, novae, supernovae etc.) of the Galaxy. In particular, we use the young stellar disk model from Refs.~\cite{Robin_2004, Knoedlseder_2005}, where the density of these sources is parameterized as
\begin{equation}
    \dot{n} = \dot{n_0} \left( e^{-(a/R_0)^2} - e^{-(a/R_i)^2} \right) \,,
    \label{eq:Disk_Model}
\end{equation}
where $a^2 = x^2 + y^2 + z^2/\epsilon^2$. We set the scale radius $R_0$ and $R_i$ to be $5$ and $3$~kpc, respectively~\cite{Vincent:2012an} and fix $\epsilon = 0.014$. This contribution is normalized to reproduce the high-longitude data points of SPI, through a simple $\chi^2$ fit to the longitude profile of the 511 keV emission at longitudes higher than $10^{\circ}$ (see more details in Sect.~\ref{sec:Result_line}).

\subsection{Continuum emission}
\label{sec:cont}

There are other emission mechanisms associated with DM production of positrons. They induce the production of a continuum $\gamma$-ray spectrum that can be used to constrain our injection model.
In addition, astrophysical sources also contribute to the emission of a continuum of $\gamma$-rays. We detail the calculation for each contribution in the following, and display them together with data in Fig.~\ref{fig:Continuum}.
 
\textbf{Ortho-positronium (o-ps) emission.}
We employ the spectral function derived by Ore and Powell 1949~\cite{OrePowell}, which is also used by the SPI publications. The o-positronium emission spectrum then reads

\begin{align}
\frac{d\phi^{o-ps}}{dE_{\gamma}} = K \frac{2}{m_e (\pi^2 - 9)} \Bigg[ & \frac{E (m_e - E)}{(2m_e - E)^2} + \frac{2m_e (m_e - E)}{E^2} \log \left( \frac{m_e - E}{m_e} \right) \nonumber \\
& - \frac{2m_e (m_e - E)^2}{(2m_e - E)^3} \log \left( \frac{m_e - E}{m_e} \right) + \frac{2m_e - E}{E} \Bigg] \,,
\end{align}
with $m_e=511$~keV being the mass of the electron in natural units.
It is normalized (i.e. we set $K$) such that the integrated emission in a given region satisfies $\phi^{o-ps}/\phi^{511} = 3.95$, with $\phi^{o-ps} = \int \frac{d\phi^{o-ps}}{dE_{\gamma}} dE'$, as was found from the analysis of Ref.~\cite{Jean_2005}. Theoretically, the value of this ratio can be $4.5$ at most, but can also take lower values down to 0.
Among all the continuum emissions from DM, the one from orthopositronium dominates at photon energies below 511~keV.

\textbf{In-flight annihilation emission.}
We use the same prescription as in Beacom and Yuksel~\cite{Beacom:2005qv} with a modification to account for the fact that injected positrons are not monoenergetic, instead having a distribution of energies (as in Ref.~\cite{Siegert:2021upf}). First, we calculate the average $511$~keV flux per solid angle, as detailed above, in the region of the SPI continuum data that we use ($|b|<15^{\circ}$, $|l|<30^{\circ}$) and then we use the following expression to evaluate the $\gamma$-ray flux from in-flight annihilations~\cite{DelaTorreLuque:2024zsr}. \footnote{This implies a simplifying assumption that in-flight annihilation occurs, on average, within the same region as the thermalised positrons. It is important to note that in-flight annihilations may, in reality, occur in different regions depending on the propagation of positrons. A more accurate estimation would require a detailed propagation model, but for the sake of simplicity, we proceed with this assumption.}

\begin{equation}
    \begin{split}
        & \frac{d\phi^\textrm{IA}}{d\Omega \,dE_{\gamma}} = \frac{d\phi^{511}}{d\Omega} \frac{n_H}{P(1-\frac{3}{4}f)} \int^{E_\textrm{max}}_{E_{\gamma}} dE'\,\frac{1}{N_{\rm pos}} \frac{dN_{\rm pos}}{dE'} \int^{E'} _{m_e}P_{E'\rightarrow E}\, \frac{d\sigma}{dE_{\gamma}} \frac{dE}{|dE/dx|} \,,
    \end{split}
    \label{eq:IA}
\end{equation}
where the second integral over $E'$ accounts
for the energy-distribution of injected positrons determined by $\frac{dN_\textrm{pos}}{dE'}$, being bounded between the highest energy used in the positron propagation simulation ($E_\textrm{max} = 5$~GeV), and the minimum positron energy able
to produce $\gamma$-rays with energy $E_\gamma$. The lower bound for the photon energy is half the electron mass.
The term $\frac{1}{ N_\textrm{pos}}\frac{dN_\textrm{pos}}{dE'}$ represents the fraction of positrons at the energy $E'$ contributing to the 511 keV emission (in the limit where this fraction is unity, we recover the case of mono-energetic positrons generating the total 511 keV flux). This takes into account the energy loss for a positron of energy $E'$ to the final energy before annihilating; the number density of hydrogen, $n_H$, counts the number of targets on which positrons scatter and lose energy.
$P_{E'\to E}$ is the probability, per unit of energy (taken from Eq.~(4) of Ref.~\cite{Beacom:2005qv}), for a positron with initial energy $E'$ to produce a $\gamma$ ray in flight before reaching the energy $E$; similarly, $P=P_{E'\to m_{e}}$ is the probability for a positron with initial energy $E'$ to produce a photon before thermalizing, and $f=0.967\pm0.022$~\cite{Jean_2005} is the fraction of positrons annihilating through positronium states.
We notice that this expression eventually becomes independent of the gas density, $n_H$, since $|dE/dx|$ (taken to be the ionization energy losses from the DRAGON2 code, including H and He -- see Ref.~\cite{DRAGON2-1} for more details) is directly proportional to $n_H$.
However, it is important to emphasize that this proportionality does not hold universally. For instance, the cross sections for interactions with H and He
  have different dependencies and weightings compared to energy losses, and the target density for direct annihilation is typically the electron density rather than the hydrogen density. Additionally, at energies above the positronium formation threshold, the direct annihilation cross section with electrons from H or He is not well characterized, though it is known to be 1-2 orders of magnitude smaller than with free electrons at lower energies. For GeV positrons, energy losses such as Inverse Compton scattering and synchrotron radiation, which do not depend on the H and He number density, become significant and do not cancel out. Thus, the assumption of independence from $n_\textrm{H}$ is valid within the specific context of sub-MeV propagation in a single-atom medium which we consider here.

\begin{figure}[t!]
\centering
\includegraphics[width=0.5\textwidth]{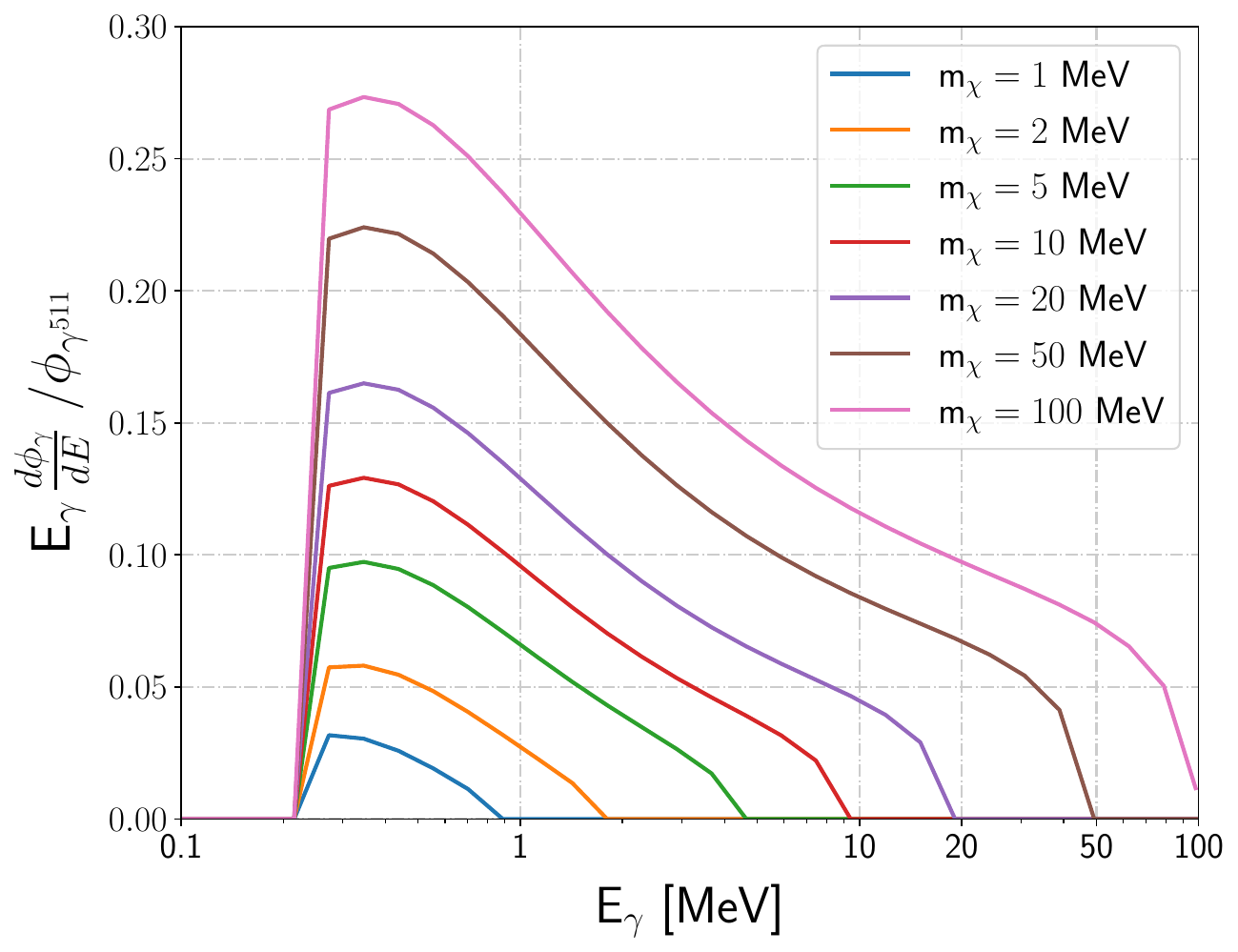}
\caption{
Flux of photons from in-flight annihilation emission of positrons, normalized to the (monoenergetic) $511$~keV flux, for DM particles with the masses indicated in the legend.}
\label{fig:IA_Rat}
\end{figure}

In Fig.~\ref{fig:IA_Rat} we show the ratio of the emission from in-flight annihilation flux over the $511$~keV flux, as a function of energy, for different DM masses spanning $1$ MeV to $500$~MeV. 
We also show the expected $\gamma$-ray flux from in-flight annihilation emission compared to SPI data in Fig.~\ref{fig:IA_Em} of App.~\ref{sec:Scaling}, where the signals are normalized to produce the total measured $511$~keV, in good agreement with the conclusions reached by Ref.~\cite{Beacom:2005qv}.
Among all the continuum emissions from DM, in-flight annihilation dominates at photon energies above 511~keV. More details about in-flight annihilation emission are provided in App.~\ref{sec:Scaling}.

\textbf{Final state radiation.}
Final state radiation (FSR) (a.k.a. inner bremsstrahlung) is included in our estimations following Ref.~\cite{Beacom:2004pe}:
\begin{equation}
    \frac{d\phi^\textrm{FSR}}{d\Omega dE_{\gamma}} = \frac{1}{2} \frac{d\phi^{511}}{d\Omega}  \frac{dN^\textrm{FSR}}{dE} \,,
\end{equation}
where
\begin{equation}
\begin{split}
    & \frac{dN^\textrm{FSR}}{dE} = \frac{1}{\sigma_\textrm{tot}}\frac{d\sigma^\textrm{FSR}}{dE}  = \frac{\alpha}{\pi} \frac{1}{E_{\gamma}} \left[\log\left(\frac{4m_{\chi}(m_{\chi} - E_{\gamma})}{m_e^2}\right)-1\right]  \left[ 1 + \left(\frac{4m_{\chi}(m_{\chi} - E_{\gamma})}{4m_{\chi}^2}\right)^2 \right]\,.
\end{split}
\end{equation}
The FSR emission is lower than the in-flight annihilation emission by a factor of a few and also lower than the o-ps emission over the relevant energies ( $>511$~keV), as expected from previous calculations (see, e.g. Ref.~\cite{sizun2007constraints}).

\textbf{Inverse Compton emission from annihilating DM.}
Once the diffuse (steady-state) distribution of electrons and positrons in the Galaxy is obtained, we make use of the {\tt HERMES} code~\cite{Dundovic:2021ryb} to integrate the CR spatial and energy distributions obtained with {\tt DRAGON2} along the line-of-sight. We use detailed interstellar gas emission maps and upto date interstellar radiation field (ISRF) models~\cite{Vernetto:2016alq} to get high-resolution sky maps of the diffuse $\gamma$-ray emission at the relevant energies. 

We calculate the ICS emission from both electrons and positrons interacting with the different ISRFs following the same procedure as in Ref.~\cite{DelaTorreLuque:2023olp}, where we refer the reader for more details. We anticipate that this emission is orders of magnitude below the other continuum emissions at MeV energies, although it becomes dominant around a few keV.

\begin{figure}[t!]
\centering
\includegraphics[width=0.6\textwidth]{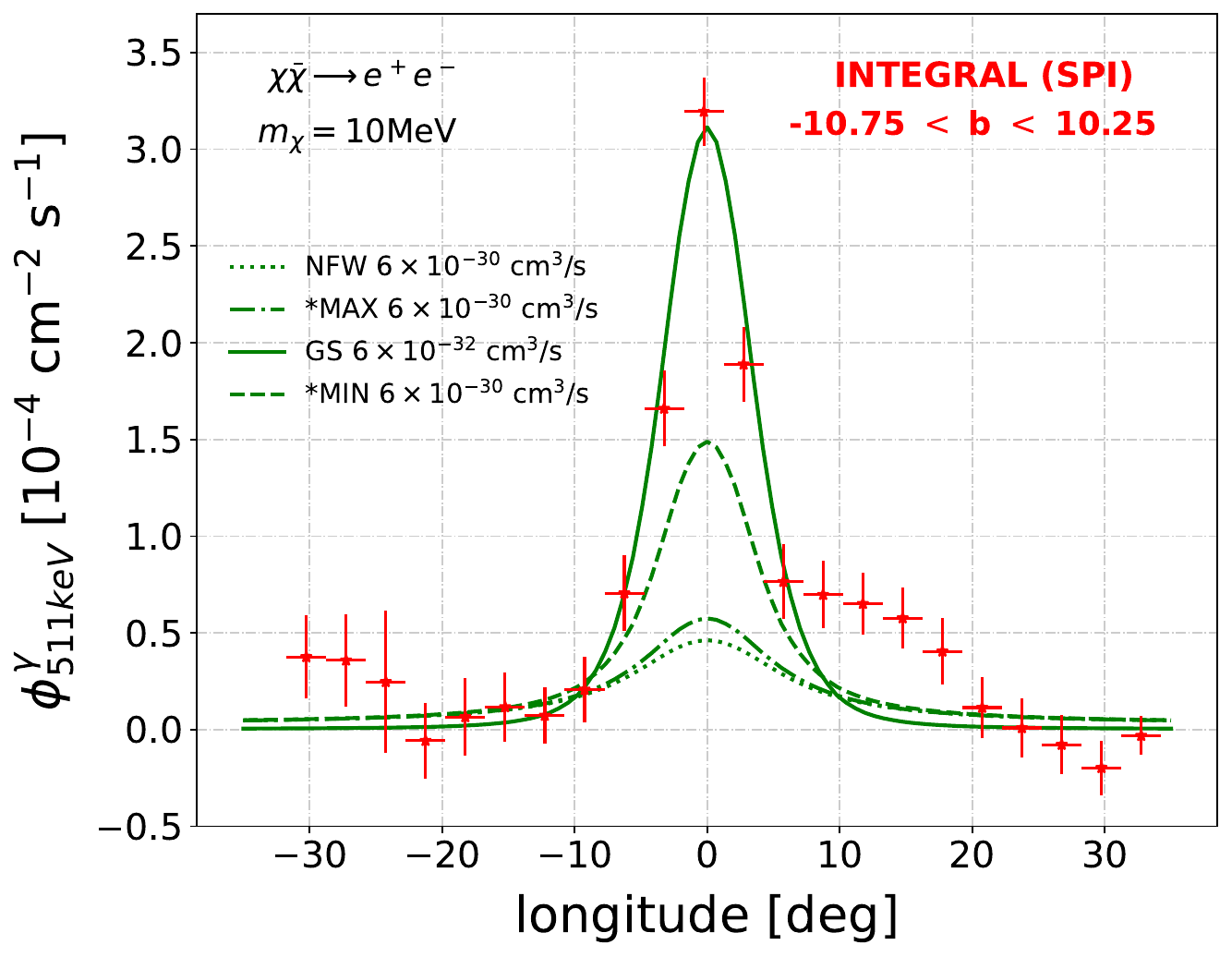}
\caption{Comparison of SPI measurements with the longitude profile of the $511$~keV line produced from DM annihilation following a NFW DM profile (dotted line), two profiles that incorporate extremal cases of stellar heating -- *MIN (dashed line) and *MAX (dot-dashed line) -- and a Gondolo-Silk (GS) profile (in the legend, as a solid line). These signals are normalized to not exceed any of the measurements at high longitudes.}
\label{fig:Comp_Profs}
\end{figure}

\textbf{Background emission.}
On top of the DM-induced $\gamma$-ray continuum signals, one has to account for two important backgrounds: the extragalactic background light and the galactic IC emission which is the dominant source here. 
For the galactic component, we use the electron model from Refs.~\cite{delaTorreLuque:2022vhm, DelaTorreLuque:2023zyd}, which is optimized to reproduce the electron and positron emission at the Earth location, as well as the local $\gamma$-ray emissivity down to a few tens of MeV. We have checked that this model shows very good agreement with the data down to $10$~keV.
The SPI instrument, which uses a coded mask for contrast measurements, is not sensitive to isotropic backgrounds. This is because isotropic backgrounds do not produce a detectable shadow pattern and are therefore degenerate with the instrumental noise.

\section{Comparison with INTEGRAL-SPI data}
\label{sec:results}
In the following section, we compare with experimental measurements of $0.1-100$ MeV photons our predictions of the signals produced from DM annihilations, for the Gondolo-Silk  *MIN, *MAX and NFW benchmarks of DM density profiles.
In particular, we use the latitudinal and longitudinal $511$~keV data from SPI~\cite{Siegert:2015knp}, as well as the continuum emission measured by SPI in the $|l|<30^{\circ}$ $|b|<15^{\circ}$ region~\cite{Bouchet:2010dj}. We also compare to COMPTEL measurements of the diffuse continuum emission~\cite{COMPTEL1994} in the same region of the sky.
From these comparisons, we show that our predicted DM signals are simultaneously compatible with the observations of the $511$~keV line and with measurements of the diffuse $\gamma$-ray emission, for DM masses of up to a $\sim20$~MeV.
A caveat that must be mentioned is that the publicly available measurements from SPI strongly depend on the templates they use to extract the observations. Therefore, these comparisons are subjected to important systematic uncertainties in the SPI measurements that we use. However, the qualitative conclusions are robust.

\subsection{511 keV line emission}
\label{sec:Result_line}

To illustrate the importance of accounting for a spike in the DM distribution, we show in Fig.~\ref{fig:Comp_Profs} a comparison of the expected longitudinal profile of the $511$~keV emission for a DM mass of $10$~MeV, for four different DM density profiles reviewed in Sec.~\ref{sec:DM}: NFW~\cite{Navarro:1995iw}, *MIN and *MAX~\cite{Balaji:2023hmy} and Gondolo-Silk~\cite{Gondolo_1999}.
In this comparison, our lines are the result of integrating the emission per unit of solid angle over $2.7^{\circ}$, which is approximately SPI's spatial resolution.
In order to be consistent with the measured disk emission (which is likely dominated by emission from astrophysical sources), these are normalized to an annihilation cross section, $\langle \sigma v \rangle$, that ensures we do not overshoot any high-longitude data point (except for the one at $30^{\circ}$, which is, in every case, exceeded by less than $2\sigma$).

One of the main takeaways from this figure is that an NFW or a relatively soft spike cannot account for a high fraction of the $511$~keV diffuse flux at central longitudes (the bulge) without contributing significantly to the disk emission. However, accounting for a DM spike can enhance the bulge-to-disk ratio, allowing for the contribution of other astrophysical sources, especially in the disk. The difference between the *MIN and the Gondolo-Silk profiles encompasses uncertainties in the spike model and show that they could suitably fit with the current observations of the $511$~keV line.

\begin{figure*}[t]
\includegraphics[width=0.496\linewidth]{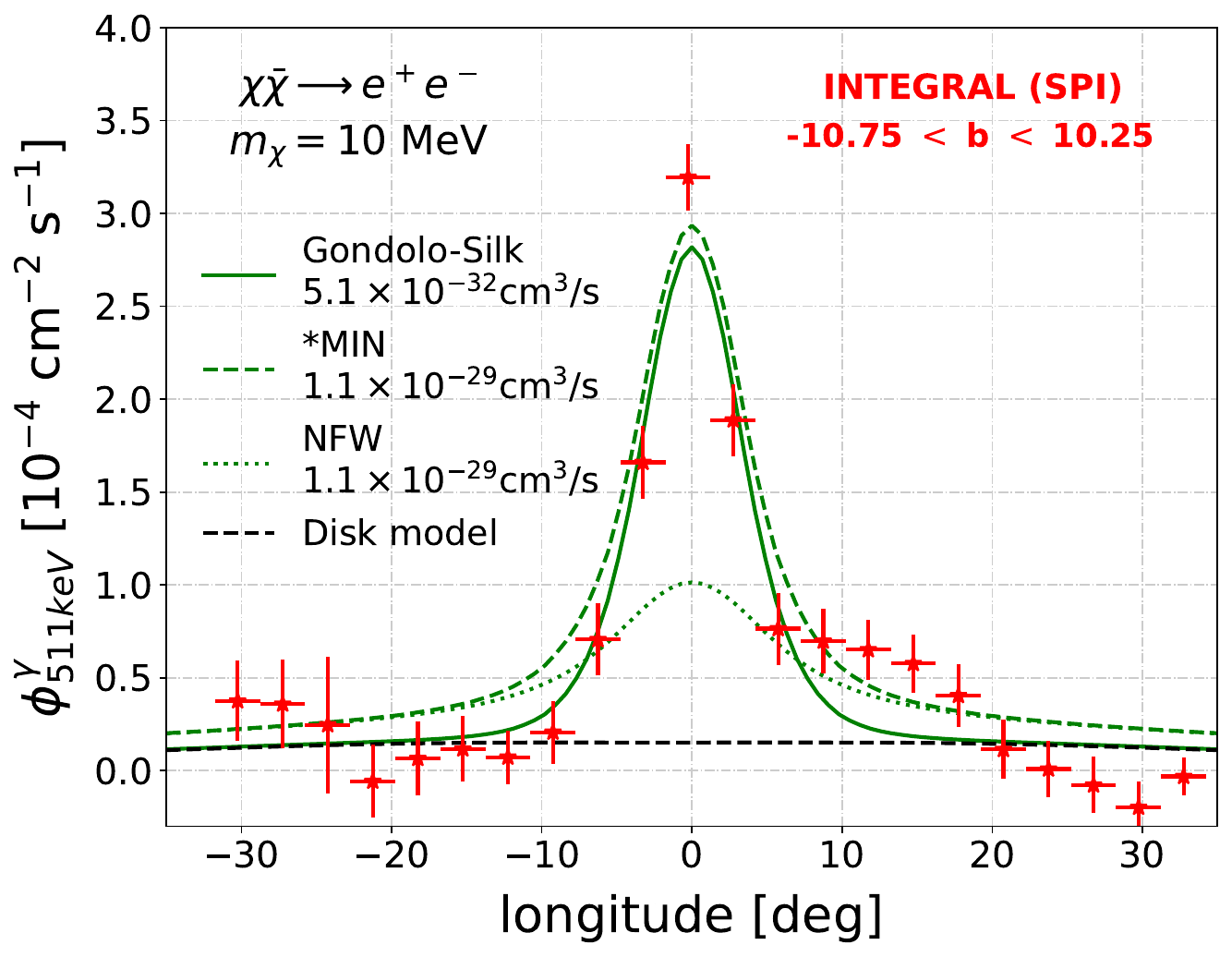}
\includegraphics[width=0.496\linewidth]{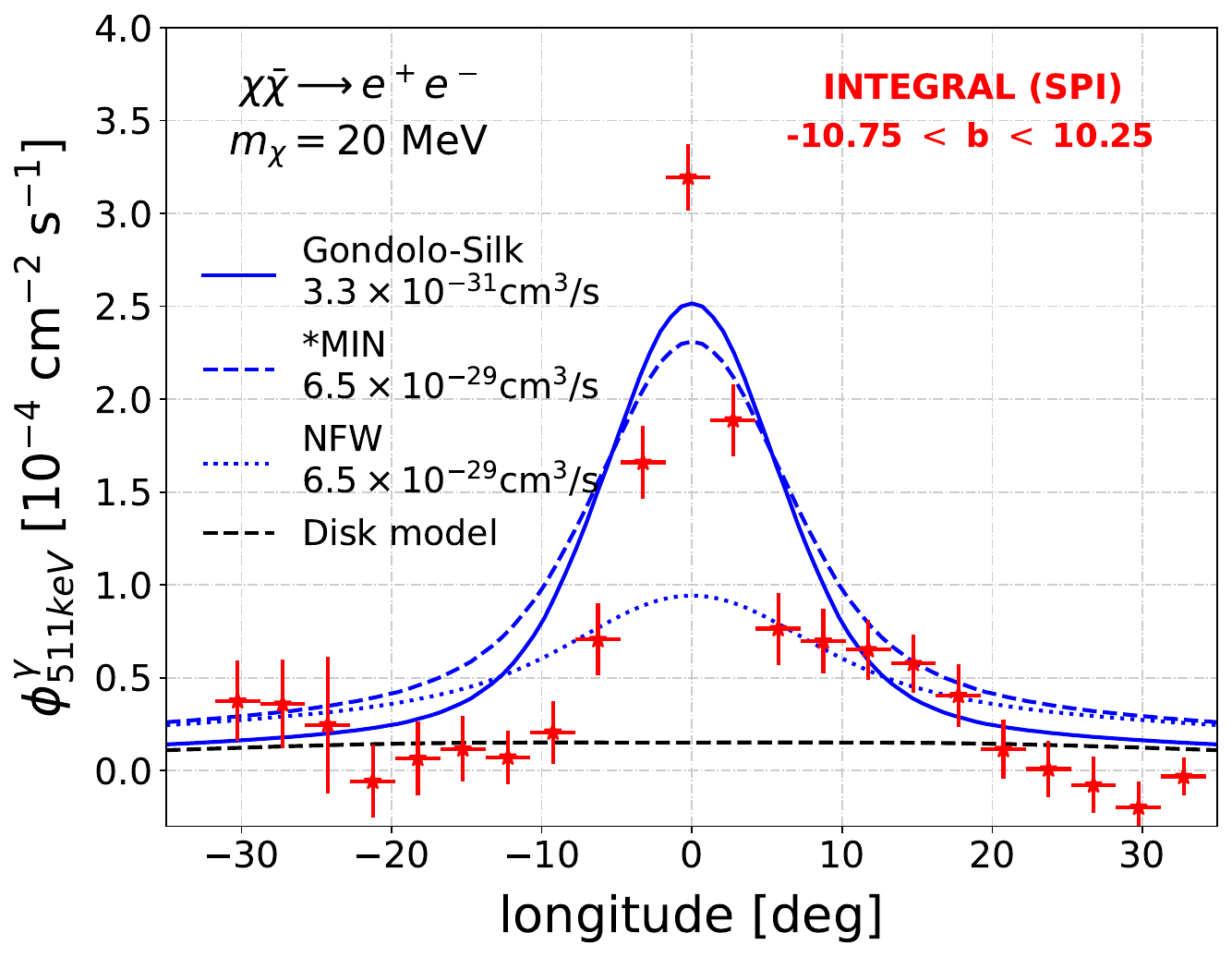}
\caption{Comparison of the best-fit signals obtained with a Gondolo-Silk and a stellar heating (*MIN in the legend) DM spike models. In the left panel we show the fits for a DM mass of $10$~MeV and in the right panel the same but for a $20$~MeV DM mass. The annihilation cross section $\langle \sigma v \rangle$ used to fit the data is shown in the legends. } 
\label{fig:GSvsSH}
\end{figure*}

In Fig.~\ref{fig:GSvsSH} we compare the $511$~keV longitudinal profile data with the fitted signal using a Gondolo-Silk, *MIN and NFW profiles for a DM mass of $10$~MeV (left panel) and $20$~MeV (right panel), including the disk emission from the young star population model explained above (see Eq.~\eqref{eq:Disk_Model}), normalized such that it does not exceed the high-longitude datapoints. 
The DM cross-sections reported in the Figure, for each DM density profile and mass, are determined from a simple $\chi^2$ test, using the \emph{curve\_fit} python package.
This comparison allows us to visualize how different the expected signals for two extreme cases of spike models can be.
We observe that in the case of $10$~MeV DM particles, there is still room for other components dominating the $511$~keV emission at mid latitudes (around $10^{\circ}$), while in the $20$~MeV case, since diffusion is faster for the higher-energy positrons injected, the profiles of emitted 511~keV photons are consequently more spread out. Even in the $20$~MeV case, an astrophysical disk plus a DM spike model can explain the data satisfactorily.

Moreover, Ref.~\cite{Siegert:2015knp} found that ``when the disk is separated into an eastern and a western hemisphere, the line widths from positive and negative longitudes also show a discrepancy at the $2\sigma$ level''.
At the moment, it seems difficult to ascertain whether this asymmetry is just an artifact coming from systematic sources of uncertainty or it has a physical origin.
However, asymmetries in the fluxes seem to be not significant, as Ref.~\cite{Bouchet:2010dj} reported.
Current analyses seem to prefer symmetric fluxes~\cite{Siegert:2015knp} when templates with a line that is slightly displaced from the GC is added (see discussions in Ref.~\cite{Weidenspointner:2008zz, Bouchet:2010dj, Skinner, Higdon_2009, Skinner:2013lla}).
In any case, it would be challenging to explain, via DM, this asymmetry or displacement of the peak emission with respect to the GC. Even if some hydrodynamic simulations of Milky-Way-like galaxies find offsets between the central DM peak and the GC (see e.g. Ref.~\cite{kuhlen2013off}), they come with a flat DM central distribution and/or could affect the DM spike. Further study is then needed to determine whether or not they could explain the asymmetric emission in our context.
A population of sources located such that they follow the spiral arms could produce an asymmetric disk emission (see Fig. 6 of Ref.~\cite{DelaTorreLuque:2023huu}), however, given the low significance of the asymmetry, we do not perform any detailed analysis of this for now. We note that the stellar bulge template use in Ref.~\cite{Siegert:2021trw} would predict such a feature too.

\begin{figure}[t!]
\centering
\includegraphics[width=7.5cm]{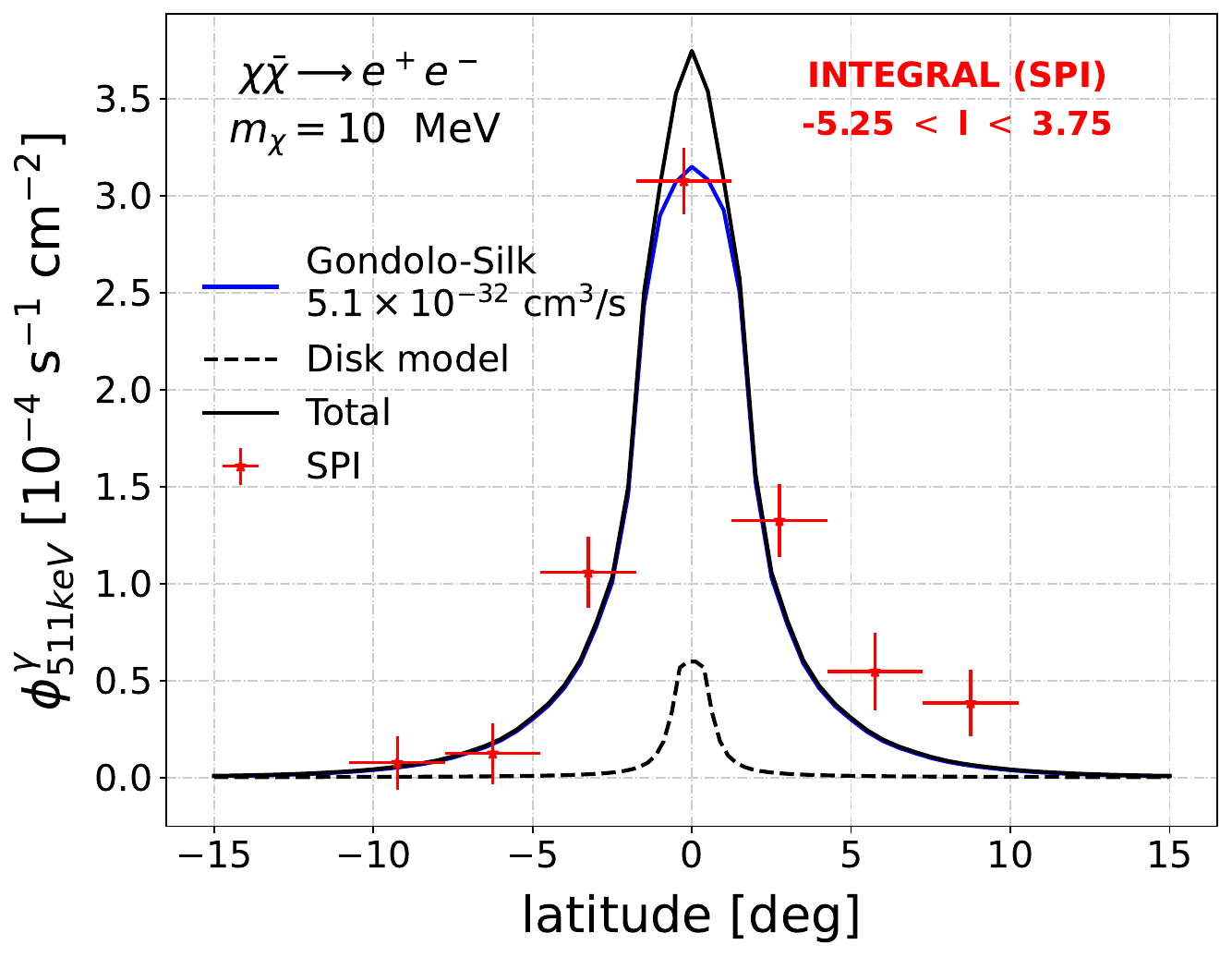}
\hspace{0.2cm}
\includegraphics[width=7.5cm]{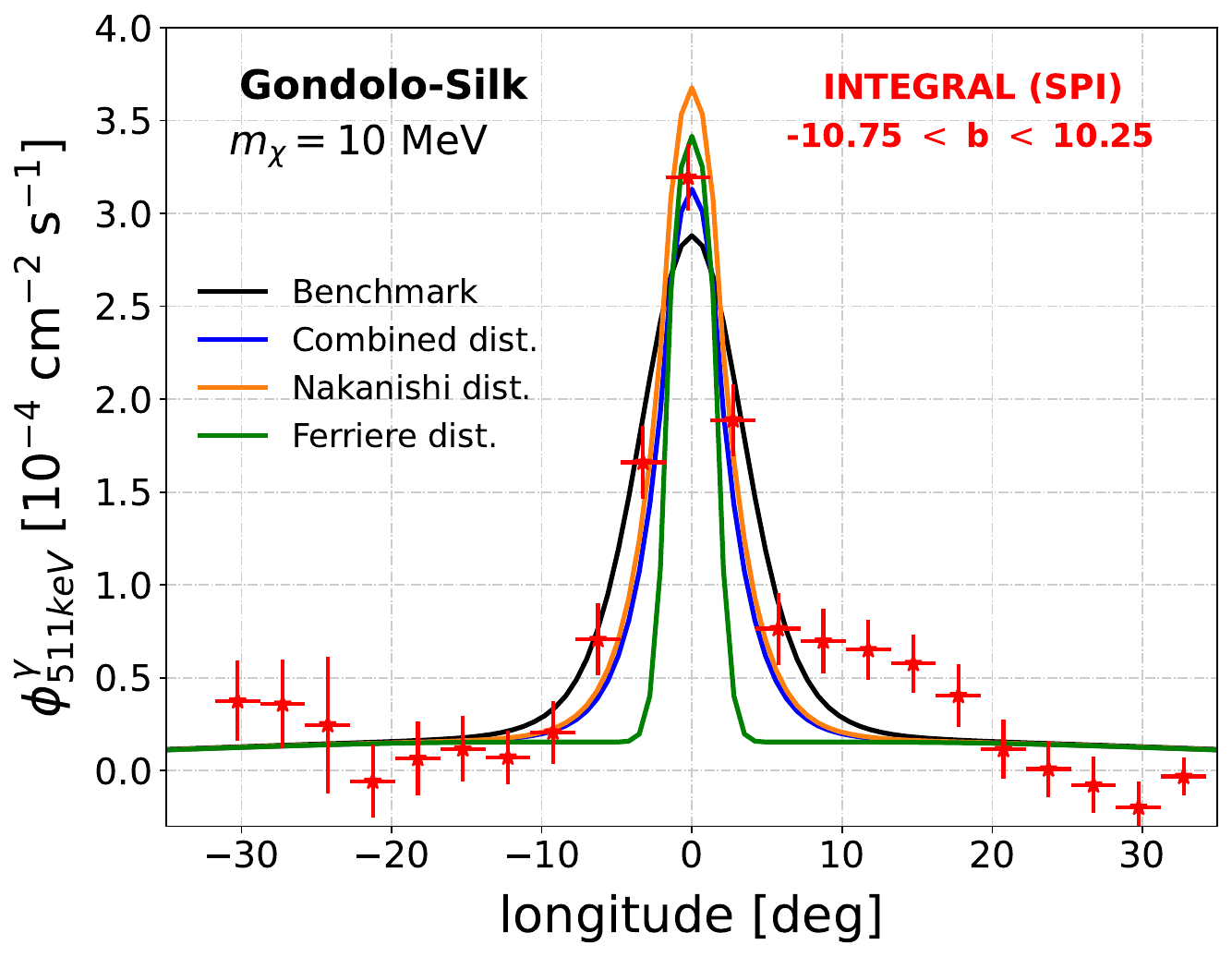}
\caption{\textbf{Left panel:} Contribution of the latitude profile of the emission at $511$~keV for the disk and DM signal (with a Gondolo-Silk profile and using a $10$~MeV DM mass) compared to SPI data. \textbf{Right panel:} Longitudinal profile of the 511 keV emission expected from DM (assuming a Gondolo-Silk profile) convolved with different gas radial distributions: The Nakanishi~\cite{Nakanishi_2003}, Ferriere~\cite{Ferriere_1998} and an even combination of Nakanishi and Ferrire distributions (``Combined dist'' in the legend).
Our benchmark assumption corresponds to a constant $n_e = 1$~cm$^{-3}$ in the Galactic plane, that makes the longitude profile follow directly the distribution of thermal positrons (see text and App.~\ref{sec:App_Uncerts} for more details).
In both panels we adopt an annihilation cross section of $\langle\sigma v \rangle = 5.1\times10^{-32}$~cm$^3$/s.}
\label{fig:GS_Profs}
\end{figure}

For completeness, we also show in the left panel of Fig.~\ref{fig:GS_Profs} the contribution of the disk and DM signals separately, for the latitudinal $511$~keV emission profile. As we see, the contribution of the disk to the central latitudes can be as important as a few tens of percent of the total measured emission. This leads to the fact that including the disk contribution should reduce the DM-induced signals (both, line-like and continuum) and relax previous constraints on DM models trying to explain the bulge emission.
Concretely, with our best-fit models shown in Fig.~\ref{fig:GSvsSH} we obtain that the disk emission have a $11.5\%$ contribution to the total $511$~keV emission in the inner $8^{\circ}$ ($20\%$ in the inner $15^{\circ}$) for the best-fit Gondolo-Silk signal at $10$~MeV mass, while for a $20$~MeV mass this contribution is of $9.5\%$ in the inner $8^{\circ}$ ($15\%$ in the inner $15^{\circ}$). Similarly,  for the *MIN profile, the disk contribution is of $10\%$ in the inner $8^{\circ}$ ($16\%$ in the inner $15^{\circ}$) for a $10$~MeV DM mass and of $9.2\%$ in the inner $8^{\circ}$ ($14\%$ in the inner $15^{\circ}$) for a $20$~MeV mass.
Additionally, in the right panel of Fig.~\ref{fig:GS_Profs}, we show the longitudinal profile of the 511 keV emission expected from DM (assuming a Gondolo-Silk profile) convolved with different radial gas distributions: The Nakanishi~\cite{Nakanishi_2003}, Ferriere~\cite{Ferriere_1998} and an even combination of Nakanishi and Ferrire distributions (``Combined dist'' in the legend). As explained in Sect~\ref{sec:line}, our benchmark case follows a constant gas density in the Galactic plane, convolved with the vertical distribution of electrons expected from the combination of the Nakanishi and Ferriere models. Therefore, the longitudinal profile in our benchmark scenario directly features the distribution of thermal positrons in the disk. See more details in Appendix~\ref{sec:App_Uncerts}.

\subsection{Continuum emission}
\label{sec:Result_Cont}

From the figures shown above, we see that DM can be the dominant source of the $511$~keV emission in the bulge, however, we require additional sources on top of it to explain the full morphology of the signal along the disk. This makes it possible for this DM candidate to have masses of up to a few tens of MeV, as long as the non-DM components do not inject high-energy positrons conflicting with in-flight annihilation emission~\cite{Beacom:2005qv} or final state radiation (FSR) constraints~\cite{Beacom:2004pe}.

\begin{figure*}[t!]
\includegraphics[width=0.496\linewidth, height=0.26\textheight]{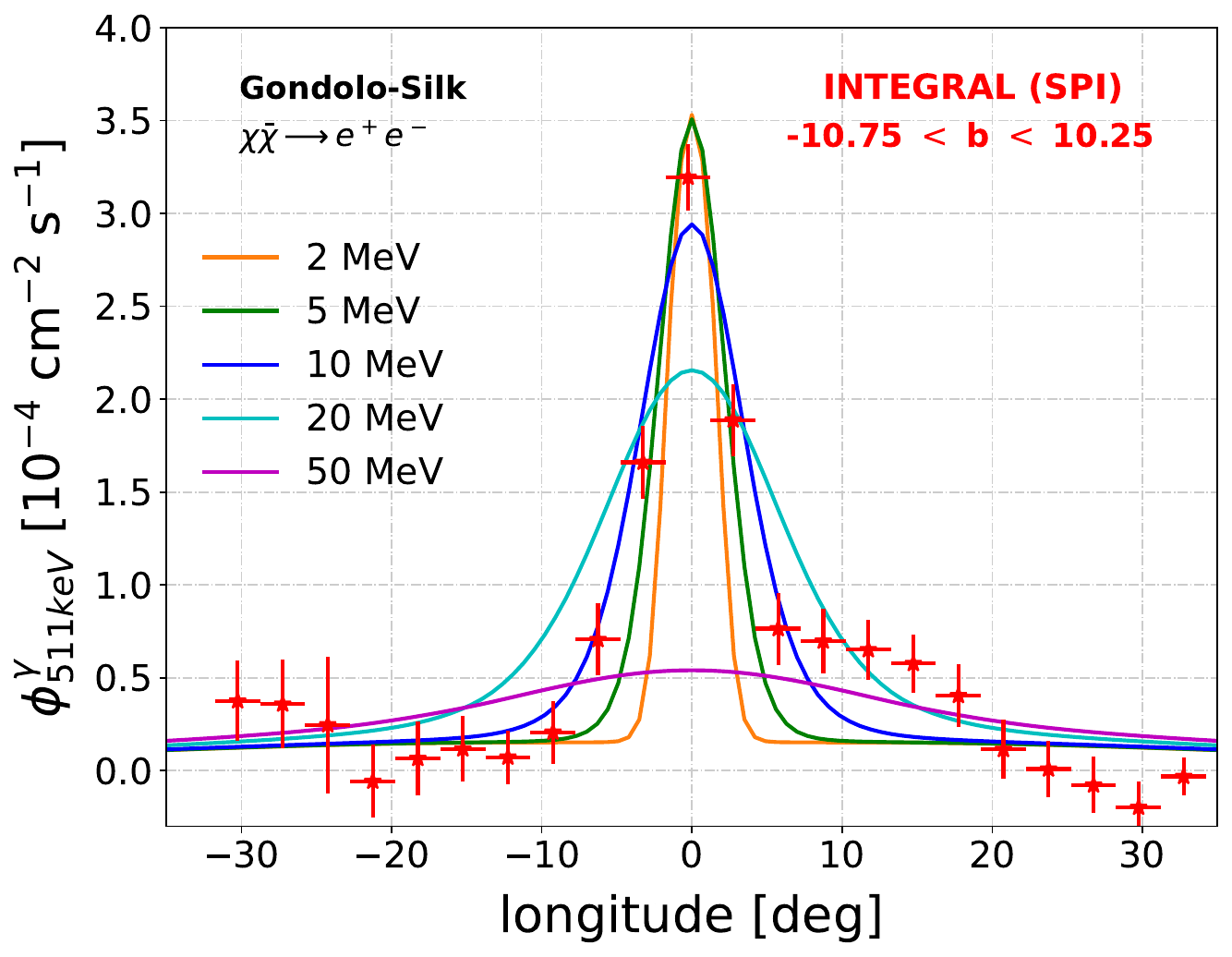}
\includegraphics[width=0.496\linewidth, height=0.26\textheight]{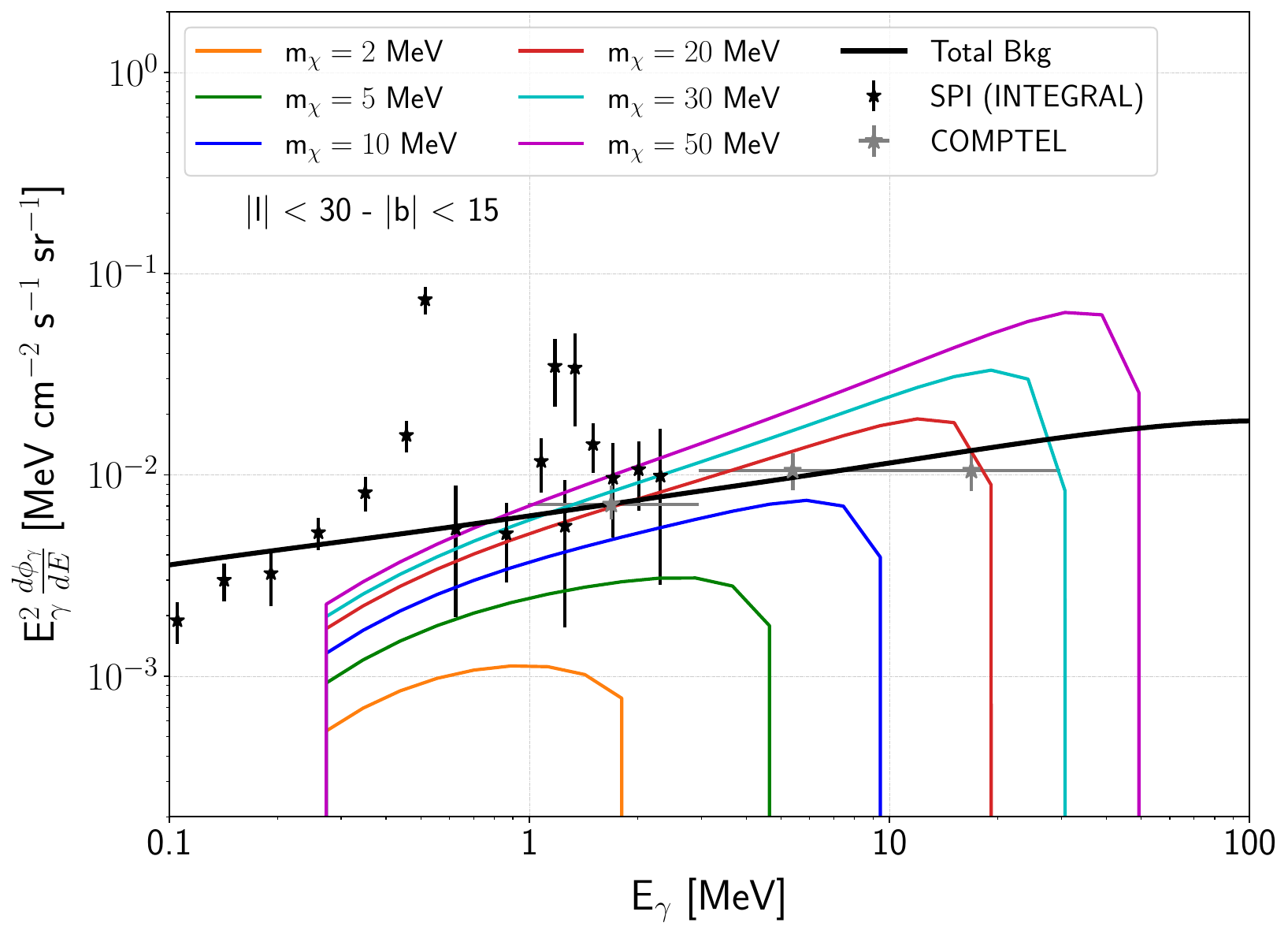}

\includegraphics[width=0.496\linewidth, height=0.26\textheight]{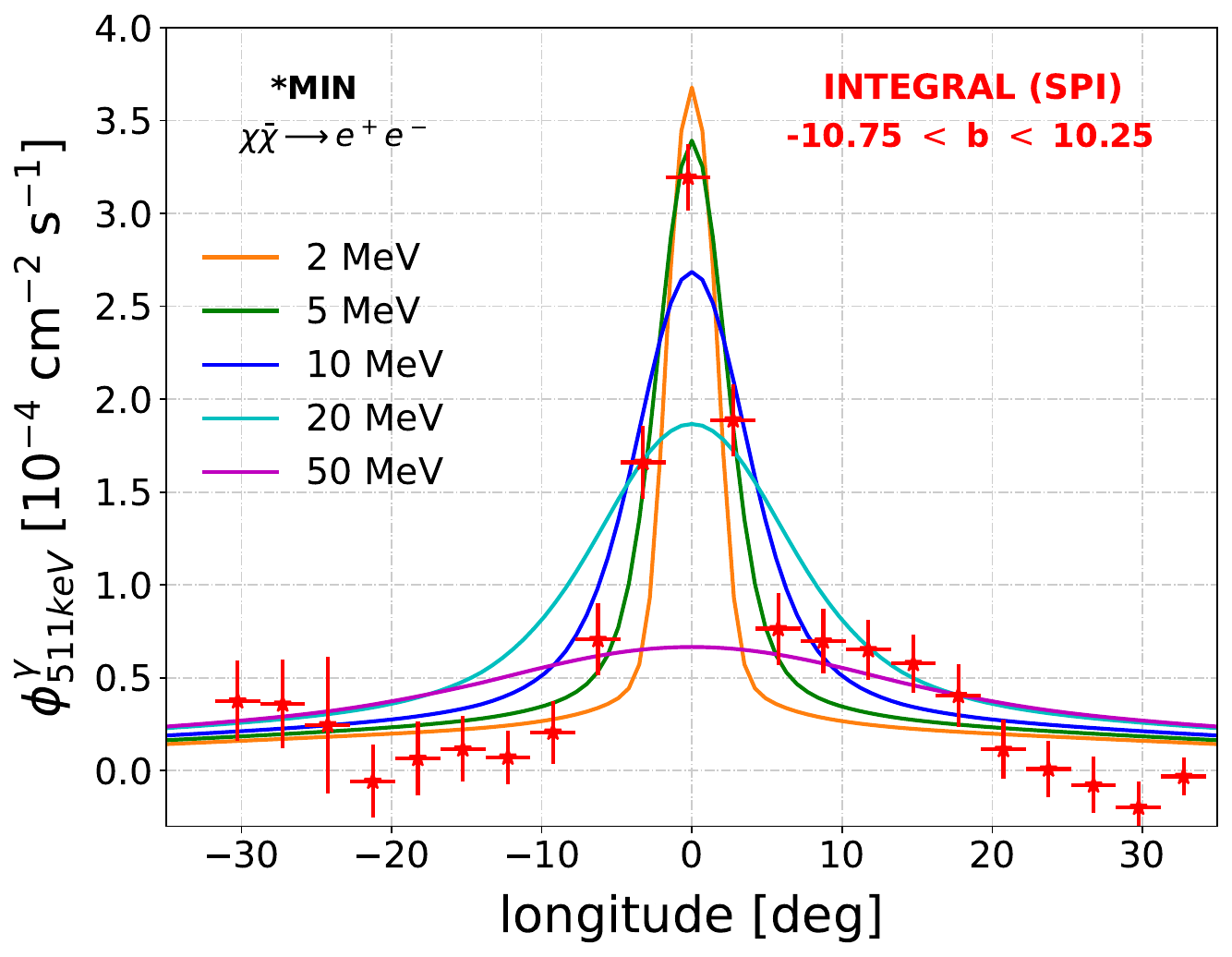}
\includegraphics[width=0.496\linewidth, height=0.26\textheight]{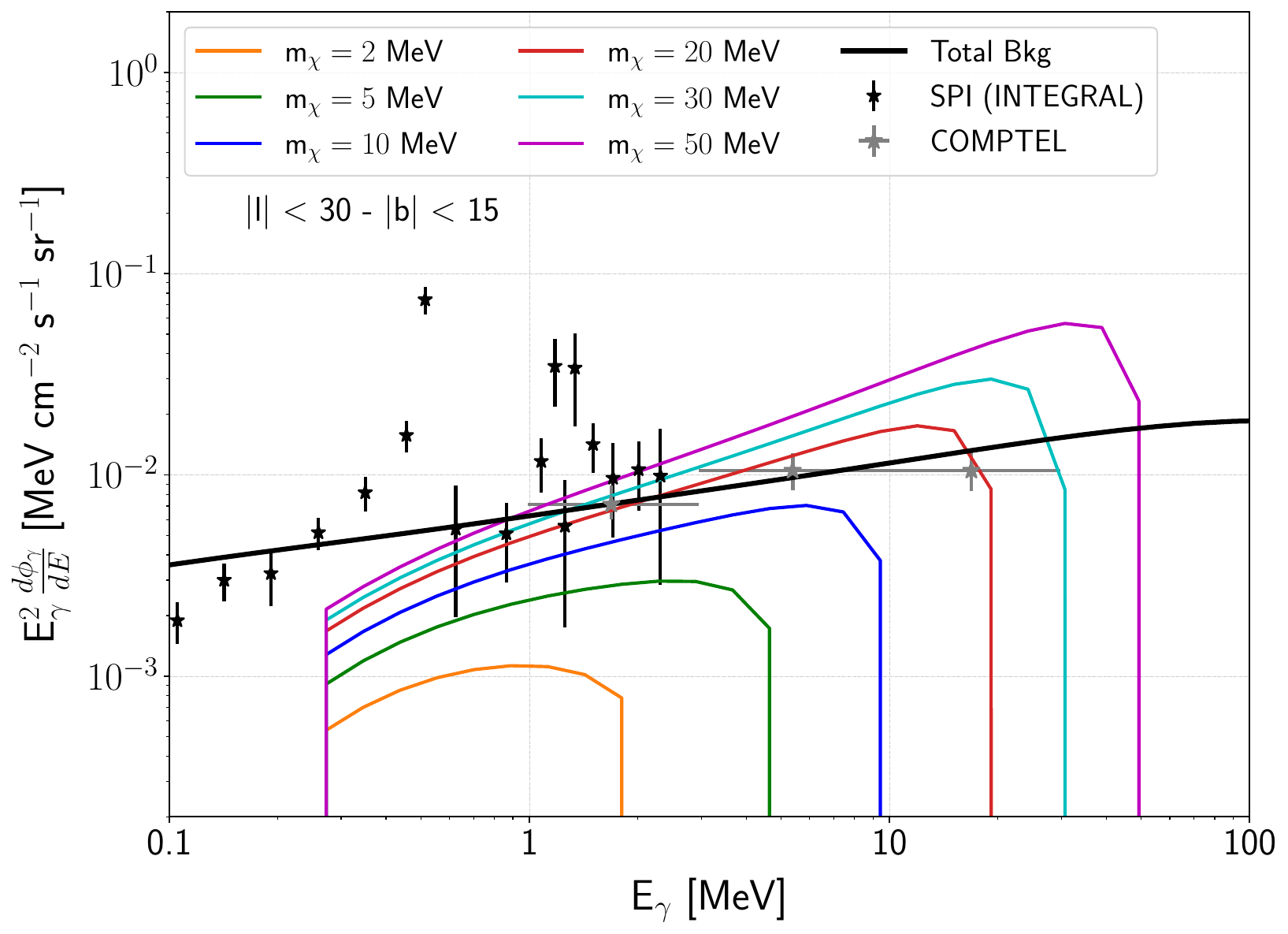}
\caption{\textbf{Left panels:} Longitude profile of the $511$~keV line as measured by SPI (red points), compared with our best-fit DM signals predicted for different masses, between $2$ and $50$~MeV, for the Gondolo-Silk (top row) and the *MIN (bottom row) DM spike profiles. \textbf{Right panels:} Predicted in-flight annihilation emission from the signals fitted to the line profile (left panels), compared to the diffuse $\gamma$-ray measurements by SPI and COMPTEL in the $|b|<15^{\circ}$, $|l|<30^{\circ}$ region. In addition, the expected background contribution is represented as a black line. The same colors representing the DM mass for each signal is  used in the left and right panels. The analogous panels for an NFW profile are dislayed in Fig.~\ref{fig:NFWContinuum}.}
\label{fig:Continuum_GS_SH}
\end{figure*}

This is evidenced from Fig.~\ref{fig:Continuum_GS_SH}, where we compare, in the right panels, the predicted in-flight annihilation emission for DM masses of  $2$ to $50$~MeV, with SPI and COMPTEL data in the $|l|<30^{\circ}$ $|b|<15^{\circ}$ region of the sky, for a Gondolo-Silk (top row) and a *MIN (bottom row) DM spike benchmarks.
For each DM mass, we perform a $\chi^2$ fit to the $511$~keV longitudinal profile (shown in the left panels) and normalize the $\langle\sigma v\rangle$ value to the best-fit, including the astrophysical disk component from stars.
The expected IC background emission is shown as a solid black line. 

From Fig.~\ref{fig:Continuum_GS_SH}, one can clearly see that masses of up to $\sim20$~MeV seem to be well compatible with the current measurements of the diffuse $\gamma$-ray galactic flux, while higher masses will significantly exceed the data (especially, COMPTEL observations) and would exceed the background emission, which is not expected. A similar figure to this one is shown in Appendix A (Figure~\ref{fig:NFWContinuum}) but for the case of a NFW DM distribution without spike.

For completeness, we show in Table~\ref{tab:XSs} our best-fit $\langle \sigma v \rangle$ values used to obtain the estimations.
We note that, as the DM mass increases, the best-fit cross section does not scale simply proportionally to $m_\chi^\alpha$ with $\alpha=2$, as one would expect simply by the DM number density and as found indeed by previous fits that ignored positron propagation (see e.g.~\cite{Vincent:2012an}). We instead find $\alpha >2$, that can be understood as follows: our signal is proportional to the number of thermalized positrons, and the larger the energy injection of the positrons (i.e. the DM mass) the less the amount of positrons that reach thermal energies before leaving the central regions of the Galaxy. Therefore larger DM masses need a slightly larger cross-section with respect to fits that ignore propagation.
Finally on the cross-sections, building particle DM models that realize them goes beyond the purposes of this paper. Still, we note that these best-fit values are compatible with existing indirect detection limits (see e.g.~\cite{Cirelli:2024ssz} for a review), and that they can easily give rise to the correct DM relic abundance via thermal freeze-out, for example in the cases of $p$-wave annihilations or of co-annihilations~\cite{Ema:2020fit}.

\begin{table}[!t]
\centering
\begin{tabular}{|c|c|c|c|c|c|}
  \multicolumn{4}{c}{\hspace{0.3cm}\large} \\ \hline  & \textbf{ 2 MeV} & \hspace{0.2 cm}\textbf{5 MeV} & \hspace{0.2 cm}\textbf{10 MeV} & \hspace{0.2 cm}\textbf{20 MeV} & \hspace{0.2 cm}\textbf{50 MeV}\\ 
  \hline
\textbf{Gondolo-Silk} & $7.6 \times 10^{-34}$ & $8.7 \times 10^{-33}$ & $5.1 \times 10^{-32}$ & $3.3 \times 10^{-31}$ & $3 \times 10^{-30}$ \\ 
\textbf{*MIN} & $2.1 \times 10^{-31}$ & $2 \times 10^{-30}$ & $1.1 \times 10^{-29}$ & $6.5 \times 10^{-29}$ & $4.8 \times 10^{-28}$  \\  
\hline
\end{tabular}
\caption{Values of annihilation cross sections ($\langle \sigma v \rangle$, in units of cm$^{3}$/s)  of self-conjugate DM used to obtain the estimations  of the photon fluxes shown in Fig.~\ref{fig:Continuum_GS_SH}, for different values of the DM mass given in the top row.}
\label{tab:XSs}
\end{table}

\begin{figure*}[t]
\includegraphics[width=0.496\linewidth, height=0.26\textheight]{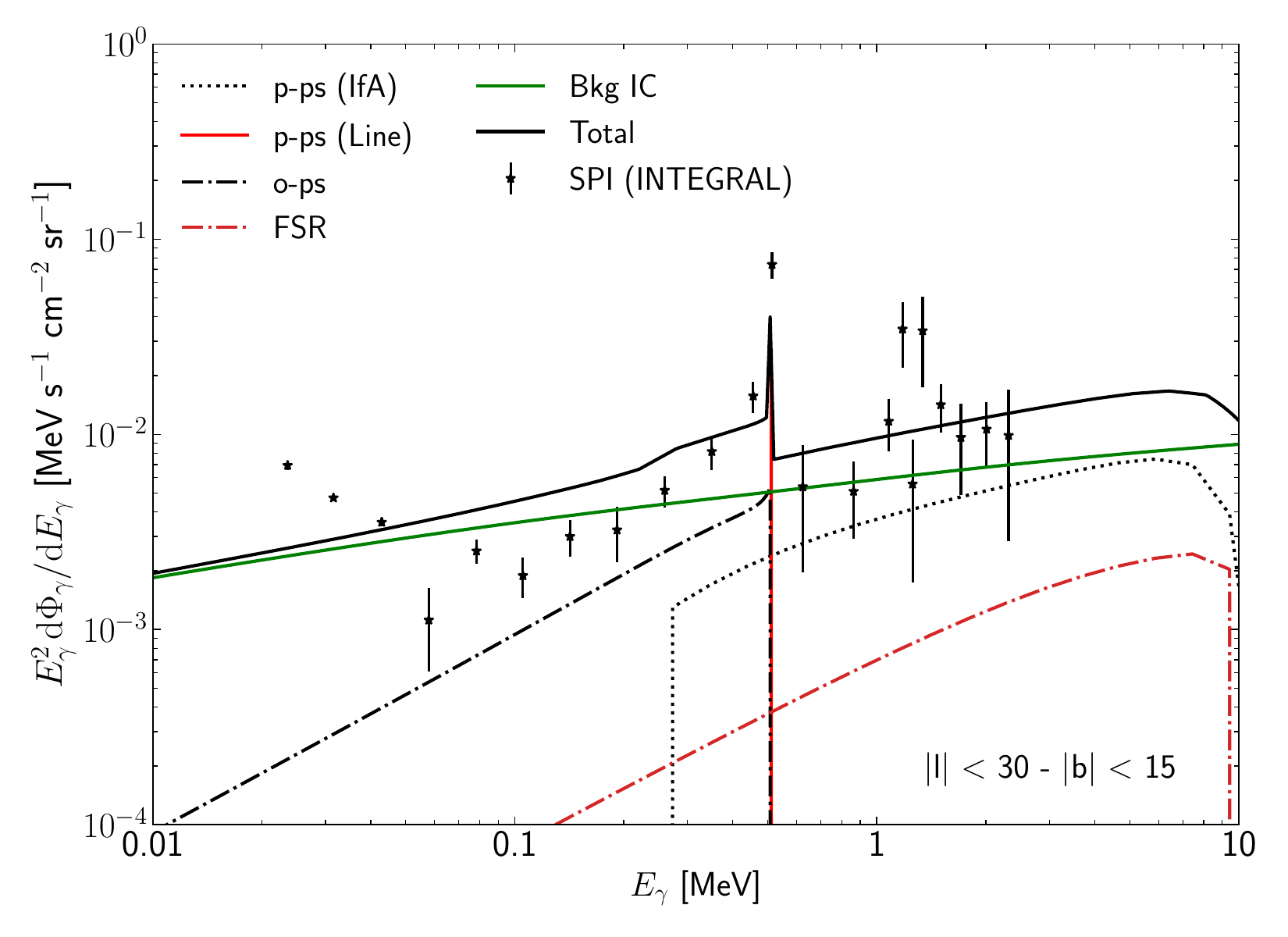}
\includegraphics[width=0.496\linewidth, height=0.26\textheight]{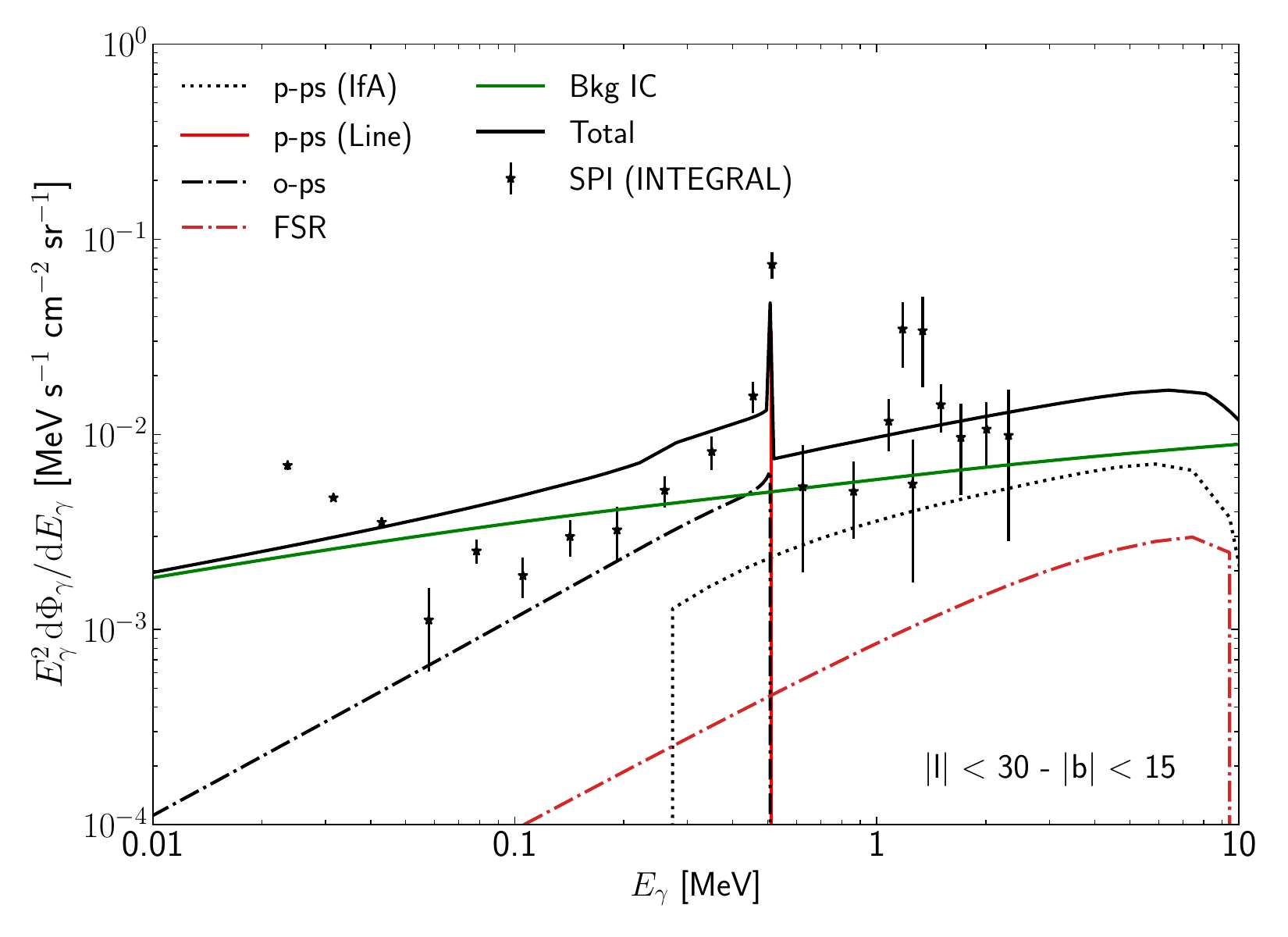}
\caption{DM-induced continuum emission components for a DM mass of $10$~MeV and the thermally averaged cross section that fits the $511$~keV line profile, for the Gondolo-Silk spike model ($\langle \sigma v \rangle = 5.1\times10^{-32}$~cm$^3$/s - left panel) and *MIN model ($\langle \sigma v \rangle = 1.1 \times 10^{-29}$~cm$^3$/s - right panel) profiles compared to SPI data. We show the o-ps (black dot-dashed lines), FSR (brown dot-dashed lines) and the in-flight annihilation (black dotted lines) emission components, as well as the expected IC background emission (green lines). The intensity of the $511$~keV line is shown as a red line. The sum of the DM induced contributions plus the background emission is depicted as a black solid line.}
\label{fig:Continuum}
\end{figure*}

Finally, in Fig.~\ref{fig:Continuum}, we show the different DM-induced continuum emission components for a DM mass of $20$~MeV and the annihilation cross-section that fits the $511$~keV line profile, for the Gondolo-Silk (left panel) and *MIN (right panel) profiles compared to SPI data. We show the o-ps (black dot-dashed lines), FSR (brown dot-dashed lines) and the in-flight annihilation (black dotted lines) emission components, besides the expected background emission (green lines). The extragalactic and IC from the injected $e^{\pm}$ by DM are not shown for clarity, since they are significantly lower than all the other components. The total emission (black solid line) sums all of these components.

We make some relevant remarks here:
\begin{itemize}
    \item For both DM spike models, as shown in Fig.~\ref{fig:Continuum}, the total predicted emission at $511$~keV is below the measured continuum emission, in the $|l|<30^{\circ}$-$|b|<15^{\circ}$ region, by at least a factor of two. This may be due to the need for other components producing additional emission at $511$~keV across this larger region, but it also could simply be due to systematic uncertainties in the data.
    \item The *MIN profile (right panel of Fig.~\ref{fig:Continuum}), leads to a slightly higher $511$~keV emission  for the best-fit annihilation cross section, since to reproduce the $511$~keV profile it needs to inject more positrons at mid-longitudes.

    \item Conservatively, i.e. without considering any background emission, the positron-induced $\gamma$-ray flux from DM is not in conflict with the data, which is remarkable given that we are considering a DM mass as high as $20$~MeV.
    \item However, including the (unavoidable) IC background emission, we observe that the low-energy (around $100$~keV) SPI datapoints may be in conflict with the total predicted flux. 
    This tension is not related to the in-flight annihilation emission induced by DM, but it seems to indicate a tension between the background emission below $511$~keV and the o-ps emission. We remark that this energy range is dominated by emission of unresolved sources. 
    In addition, we note that the o-ps emission does not depend on the DM mass, but only on the ratio of the p-ps emission to the o-ps, which we set to be $3.95$, according to the analysis of older SPI data, from Ref.~\cite{Jean_2005}. As we comment around Fig.~\ref{fig:ops}, the SPI data below $511$~keV seem to indicate that the IC background must suddenly drop exactly at $511$~keV, which is not physically motivated. Therefore, this incompatibility with the low-energy SPI datapoints seems to be due to the high systematic uncertainties not shown in the data. 
    \item Above $511$~keV, where the in-flight annihilation emission dominates, we observe that the total predicted emission does not significantly exceed the SPI data for $m_{\chi}\lesssim20$~MeV. In this energy region, the SPI observations become more sparse, mainly due to the presence of the $1.8$~MeV line from $^{26}$Al decay and the $\sim1.3$~MeV line from $^{60}$Fe decay. However, for the *MIN profile, the in-flight annihilation emission becomes more important than the galactic background, which would lead to a distinctive feature that could be observed at higher energies. 
    \item A possible test of our proposal would come from data in ROIs other than $|l|<30^\circ$-$|b|<15^\circ$, as well as from different experiments. To the best of our knowledge, these are provided by COMPTEL and EGRET, which we show and discuss in App.~\ref{sec:ROIs}, finding that they do not alter our conclusions.
    We also include similar comparisons with SPI data in a ROI of $47.5^{\circ}$ around the GC, reaching energies of up to $8$~MeV (Fig.~\ref{fig:IA47.5} in App.~C). An important point of this dataset is that systematic uncertainties below $\sim100$~keV (where IC on starlight becomes dominant) are better accounted for.
    
\end{itemize}
These remarks indicate that spiked DM profiles are not in significant tension with the current $\gamma$-ray keV-MeV data for DM masses of up to $\simeq 20$~MeV. With respect to previous analyses, our results increase by at least a factor of a few the largest mass of DM whose annihilation can explain the bulge line.

\section{Summary and discussion}
\label{sec:Conclusion}

Around $50$ years after the firm detection of a bright $\gamma$-ray line at $511$~keV from the bulge and disk of our Galaxy, its origin still remains unknown. Since a few sources of positrons are known to possibly contribute to this emission, the observations would likely be the product of their combination. However, there is no clear candidate able to explain the high flux observed from the bulge. Here, we have revisited the hypothesis of sub-GeV DM annihilations as the dominant source of $511$~keV photons from the bulge.

For the first time in this context, we have considered the possibility that the DM density profile develops a spike around Sgr A*. This is expected for peaked DM profiles like NFW, that were already known to be needed to possibly explain this signal. We have computed the associated 511~keV line emission and demonstrated that it can fit the bulge data whilst producing very little emission at the disk, which we have assumed to have originated mainly from astrophysical sources, e.g. young stars.
Remarkably, this conclusion holds not only for Gondolo-Silk spikes, but also for more conservative ones that take into account DM profile softening by stars in the vicinity of Sgr A* and are supported by simulations (all our spike benchmarks respect the limits on the mass close to Sgr A* by the GRAVITY experiment~\cite{GRAVITY:2021xju}).
DM profiles like NFW without a spike, that explained the observed 511~keV bulge emission, have been recently found to produce too much disk emission --which we confirmed here-- and hence they were not favoured by data. Our findings, therefore, motivate new interest in the DM hypothesis for the 511~keV line.
Although we remark that the measurements of the distribution of the line emission are very challenging and are subject to important systematic uncertainties, our work shows that spike profiles can reproduce the data satisfactorily adopting the positron propagation that is extrapolated from CR observations at higher energies.

Importantly, we also find that DM spikes imply that the in-flight annihilation emission, associated with the explanation of the 511~keV line in the bulge, is compatible with MeV diffuse $\gamma$-ray observations for DM masses upto $m_\chi\simeq 20$~MeV, possibly higher for the Gondolo-Silk benchmark.
This is due to the smaller DM annihilation cross sections needed to explain the bulge data (due to the higher DM density there), along with our inclusion of state-of-the-art propagation of positrons after they are injected by DM, and the inclusion of the contribution from massive stars dominating the disk emission. 
Our finding that DM masses up to $20$~MeV can explain the 511~keV line is unlike previous studies, that found that DM masses below a few MeV were needed. Those low masses were generically excluded by CMB and BBN, unless ad-hoc neutrino injections in the early universe were postulated. Our findings then remarkably simplify the task to find DM models for the 511~keV line, and open model-building avenues that await exploration.
For example, models of MeV DM for the 511~keV line predicted that they would have soon be tested by direct detection of electron recoils~\cite{Ema:2020fit}. It will be intriguing to see what our findings imply for direct detection, collider searches and other tests of the DM hypothesis.

Finally concerning our new results, we have tested how some variations, either in the diffusion setup or in the electron density distribution in the Milky Way, can affect the reported spatial profiles of the emission. As we discuss in Appendix~\ref{sec:App_Uncerts}, variations of our benchmark scenario mainly lead to a more peaked profile, which leads to an even higher ratio for the bulge-to-disk emission and thus reinforce our conclusions.
Furthermore, an inhomogeneous diffusion scenario where CRs are more confined around the GC~\cite{Luque_PeV}, as motivated by observations of the Fermi-LAT~\cite{Gaggero:2014xla, Fermi-LAT:2012edv,Fermi-LAT:2016zaq}, will also lead to a more peaked profile of the predicted line. However, winds in the GC will have the opposite effect. 

\medskip

We now move to comment on observational implications of our results for MeV telescopes. First, we note that the SPI measurements employed are significantly affected by the templates used to extract the observations. In particular, the data that we are using are extracted using Gaussian templates, which have a similar symmetry as the one expected by DM-like models. Recently, stellar templates for the nuclear stellar bulge and a boxy bulge were shown to be statistically preferred over the profile expected from an NFW DM distribution~\cite{Siegert:2021trw}, even though positron propagation was not included on top of NFW and thus, strictly speaking, the conclusion applies only to DM masses closest to an MeV. 
In addition, Ref.~\cite{Siegert:2021trw} found that the NFW template was the best model when analyzing
the $511$~keV line, while it is the continuum o-ps emission what makes their analysis favour the stellar bulge hypothesis.
Moreover, the templates of~\cite{Siegert:2021trw} did not consider the DM spike profiles that we are exploring here.
Therefore, our goal here is not to provide the best and most accurate explanation of the $511$~keV line emission, which would require access to SPI's raw data and the coded mask response function, but to demonstrate that the $511$~keV observations and the correlated in-flight annihilation emission motivate considering the possibility of an origin from DM annihilations alleviating previous constraints.

A promising way to the reveal the presence of sub-GeV DM concentrated in a spike would be to measure the continuum $\gamma$-ray diffuse emission around Sgr A* and search for bump-like structures, since the $511$~keV emission must be dominated by this $10$-$20$ MeV DM particle and the related in-flight positron annihilation emission should show clear signatures. 
In fact, the spectral analyses of Ref.~\cite{Siegert:2015knp} reveals that a central component in their fit (what they call a ``GC Source") seems consistent with Sgr A* as a source of positrons. However, given the uncertainties in the analysis, it may be too soon to claim that the vicinity of Sgr A* is a clear source of $511$~keV photons, and indeed more observations would be needed to confirm this. 
Additionally, other backgrounds may be present and features from sub-GeV DM may not be so easily observed.
Another possible smoking gun for indications of a spike around Sgr A* would be the observation of a red or blue-shift of the $511$~keV line. Possible detection of o-ps emission at energies higher than $511$ keV or in-flight annihilation emission below $\sim 100$~keV will also lead to a similar  conclusion. This is due to the fact that the spike is expected to be rotating in the same way as Sgr A*.
Finally, the existence of a spike of DM annihilating into $e^+e^-$ could possibly be tested by BH observations with the Event Horizon Telescope~\cite{Chen:2024nua}.


There are a few experiments being proposed to improve the current sensitivities in the MeV gap that may probe our hypothesis.  
These include the Compton spectrometer and imaging telescope  (COSI) with  improved sensitivity over INTEGRAL-SPI,  to be launched in 2027. 
COSI~\cite{tomsick2019compton, Siegert_2020, Karwin_2023} is expected to improve the sensitivity to  line and continuum emission over 0.2 to 10  MeV by   up to an order of magnitude.
Interestingly, this mission will be able to improve the measurements of the disk emission, which currently lead to the best constraints on the properties of different kinds of electrophilic feebly interacting particles~\cite{DelaTorreLuque:2023huu, DelaTorreLuque:2023nhh} or sub-GeV DM~\cite{DelaTorreLuque:2023cef}.
On a longer time-scale, the Galactic Explorer with a Coded Aperture Mask Compton Telescope (GECCO)~\cite{Orlando_2022, Coogan_2023}, is expected to provide measurements with very high angular resolution on arc-minute scales up to $\sim10$~MeV. Amego-X~\cite{Caputo_2022} is also expected to perform measurements in the  MeV band with unprecedented accuracy  and with significantly higher effective area than INTEGRAL-SPI and COSI.

\section*{Acknowledgements}
We thank Michael Mancini and Paolo Panci for useful comments on the manuscript.
SB and MF are supported by the STFC under grant ST/X000753/1.
PDL is supported by the Juan de la Cierva JDC2022-048916-I grant, funded by MCIU/AEI/10.13039/501100011033 European Union ``NextGenerationEU"/PRTR. The work of PDL is also supported by the grants PID2021-125331NB-I00 and CEX2020-001007-S, wich are both funded ``ERDF A way of making Europe'' and by MCIN/AEI/10.13039/501100011033. PDL also acknowledges the MultiDark Network, ref. RED2022-134411-T. This project used computing resources from the Swedish National Infrastructure for Computing (SNIC) under project Nos. 2021/3-42, 2021/6-326, 2021-1-24 and 2022/3-27 partially funded by the Swedish Research Council through grant no. 2018-05973.
FS is partly supported by the Italian INFN program on Theoretical Astroparticle
Physics (TAsP), by the French CNRS grant IEA ``DaCo: Dark Connections'' and by COST
(European Cooperation in Science and Technology) via the COST Action COSMIC WISPers
CA21106.


\bibliographystyle{JHEP}
\bibliography{references}

\appendix

\section{Uncertainties in the predicted profiles and mean distance travelled by positrons}
\label{sec:App_Uncerts}

\begin{figure*}[b!]
\centering
\includegraphics[width=0.65\linewidth]{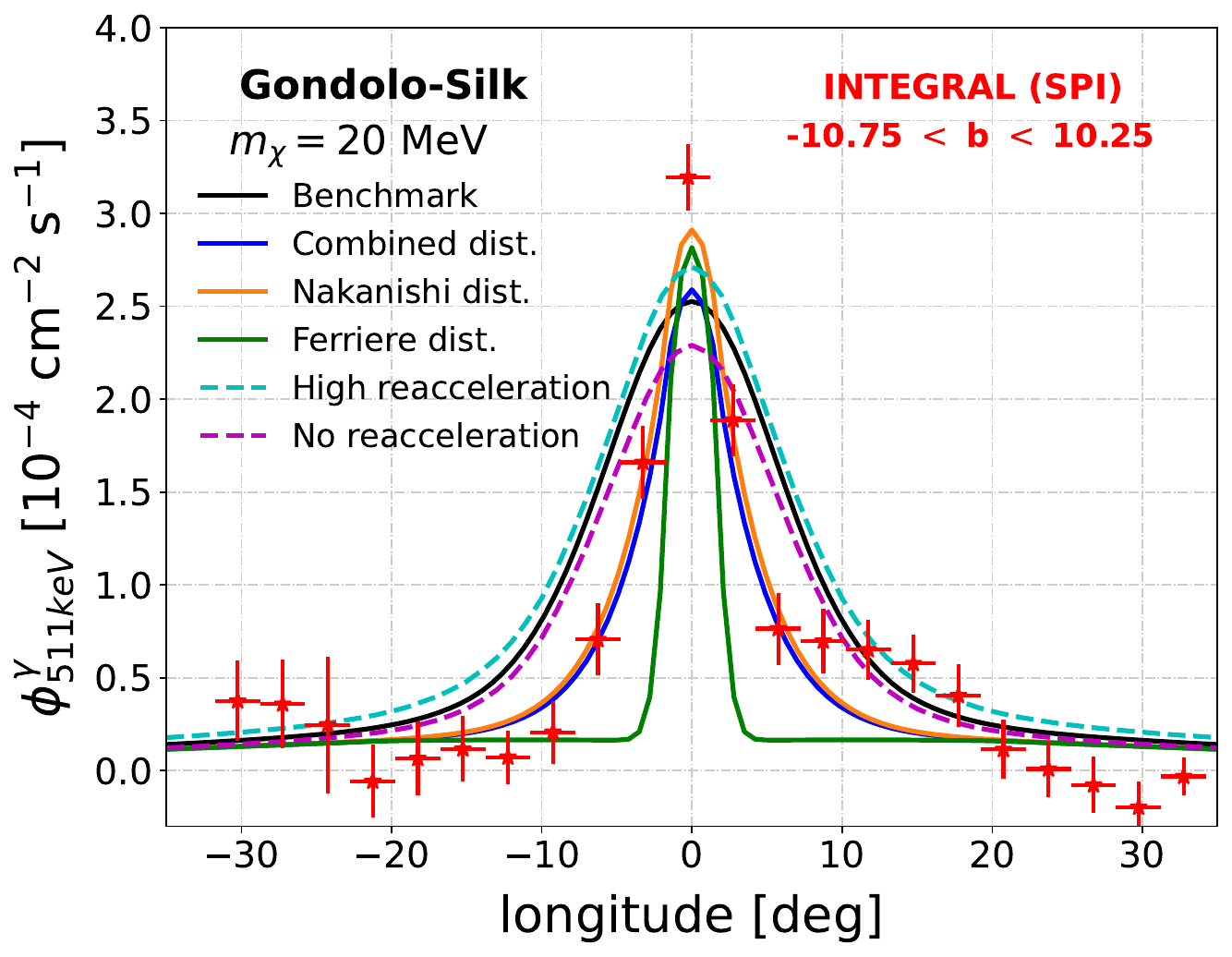}
\caption{Effect of different assumptions in the predicted $511$~keV emission longitude profile. The dashed lines represent the cases where we vary between two extreme benchmark values for the Alfven speed (that regulates the level of diffuse reacceleration that particles undergo). The solid lines show the predicted profiles adopting two different radial distributions for the electron density. All these cases are evaluated with a Gondolo-Silk spike model, for a mass of $m_{\chi} =20$~MeV and with cross sections $\langle \sigma v \rangle = 3.3\times10^{-31}$~cm$^3$/s).}
\label{fig:Uncerts}
\end{figure*}

In this appendix, we discuss how different assumptions can change our predicted line emission. 
First, regarding uncertainties in the spatial distribution of diffuse positrons, we note that the main relevant variables are the propagation parameters, the gas distribution and the injection of the positrons. The injection of the positrons follows the spike models described above, for which the normalization is controlled by the DM density at Earth and annihilation cross section $\langle \sigma v \rangle$, for a given DM mass. These parameters do not affect the profile of the predicted $511$~keV line, only the normalization of the emission. The gas distribution would affect the energy losses of positrons, which is the dominant process at DM masses below a few tens of MeV. 

In our simulations, we employ the gas maps developed by the GALPROP team~\cite{Moskalenko:2001ya, Ackermann:2012pya}, which is one of the most popular gas distributions employed in CR propagation studies. A different gas distribution is not expected to change our conclusions significantly, but it may affect our predictions for the zones very close to the GC, since different gas distributions can vary noticeably around the GC, often predicting a higher gas density than the one used in this work (albeit not significantly modifying the predictions shown). We note that the available gas distributions in current CR propagation codes are not optimized for very small scales (tens of parsecs) around the GC. 
Using a gas distribution with a higher density around the GC (which could happen if the region in the vicinity of Sgr A* concentrates more gas than assumed here) would lead to higher energy losses, which means that the positrons will remain closer to the DM distribution while thermalizing and, therefore, result in a more peaked emission profile and a higher bulge-to-disk ratio, which would further reinforce our conclusions.

In terms of uncertainties related to the propagation parameters, it has already been pointed out that reacceleration is the main source of uncertainty for sub-GeV DM in Ref.~\cite{DelaTorreLuque:2023olp}. We have tested the effect on the spatial distribution of the line for two extreme benchmark cases of no reacceleration ($V_A = 0$~km/s) and very high reacceleration ($V_A = 40$~km/s). Different values for the halo height will only rescale our predictions, with no relevant implications on the shape of the profile. The difference in the predicted profile of the line is small, as can be seen in Fig.~\ref{fig:Uncerts} (dashed lines) for the case of a $20$~MeV DM particle, following the Gondolo-Silk spike model. We notice that for higher masses, the effect of reacceleration is even lower.

\begin{figure*}[b!]
\centering
\includegraphics[width=0.49\linewidth]{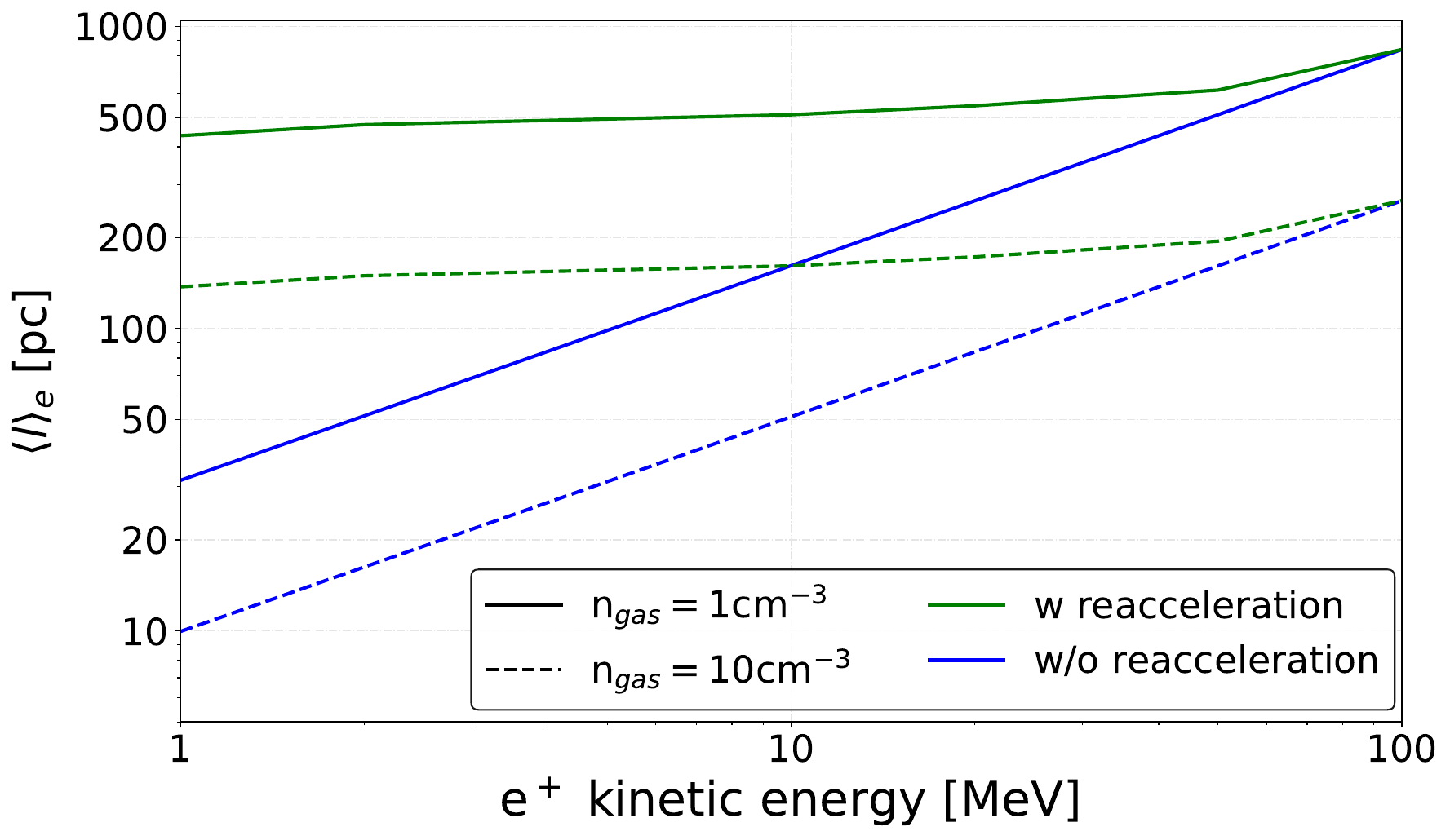}
\includegraphics[width=0.48\linewidth, height=0.195\textheight]{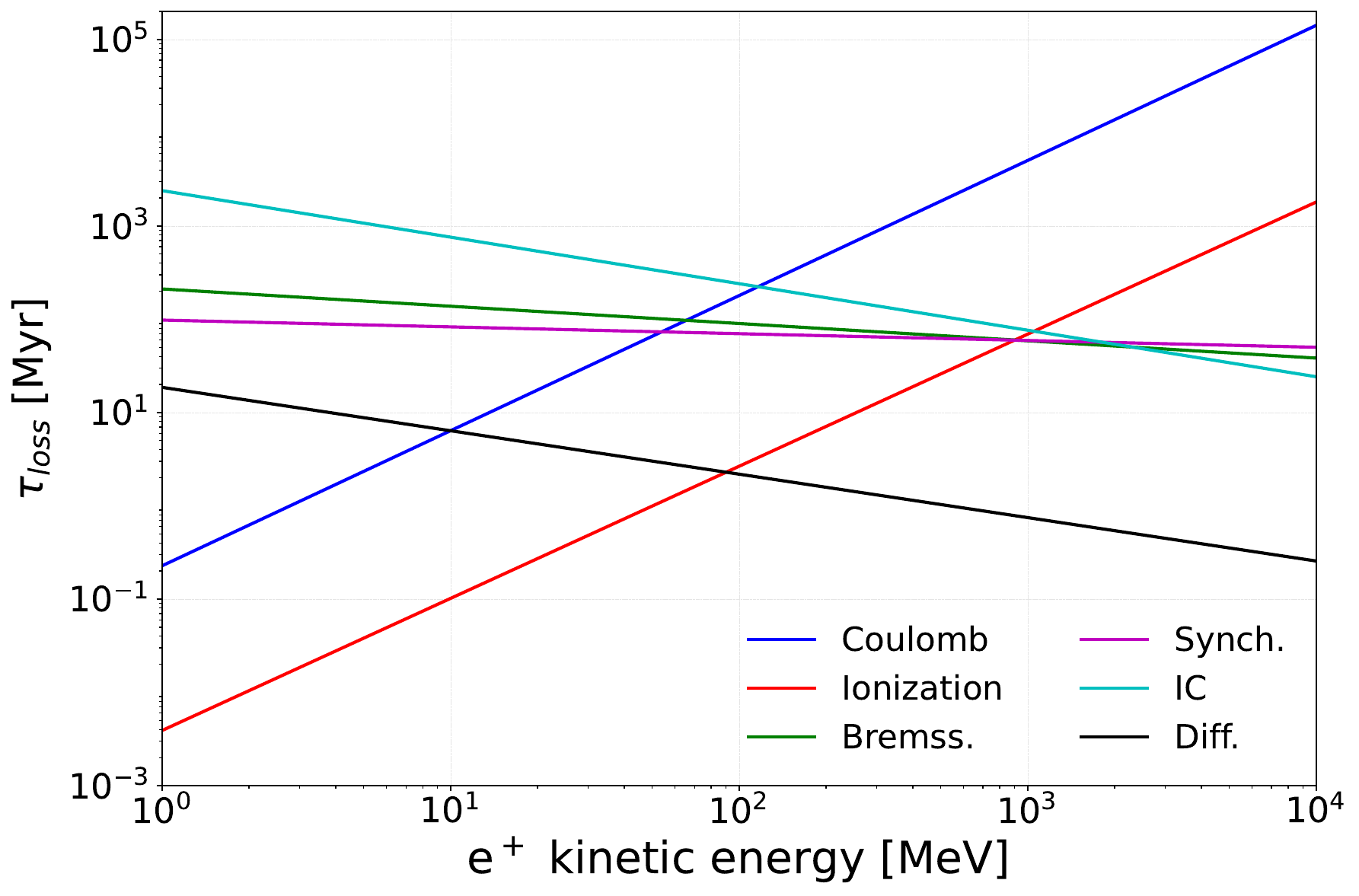}
\caption{\textbf{Left panel:} Average distance traveled by positrons in a medium with total density of $1$~cm$^{-3}$ (solid lines) and $10$~cm$^{-3}$ (dashed lines), assuming $10\%$ of He. The blue lines denote the average distance traveled by positrons without including the effect of reacceleration and the green lines represent a rough estimation of the mean distance traveled by positrons when including the effect of reacceleration. \textbf{Right panel:} Loss time for the main energy loss mechanisms for a medium of $1$~cm$^{-3}$ and a magnetic field of $5$~$\mu$G.}
\label{fig:MeanFreePath}
\end{figure*}

Second, as shown in Eq.~\eqref{eq:511line}, the spatial morphology of the line also depends on the distribution of ambient electrons available to form positronium bound states. Our benchmark predictions estimate a flat distribution of electrons along the Galactic disk and an exponential decrease away from it. We have tested how the assumption of a flat electron distribution in the Galactic disk changes our predictions. First, we have adopted the Ferriere distribution~\cite{Ferriere_1998} of ionized hydrogen as implemented in the DRAGON2 code. 
 The predicted profile is shown as a green line in Fig.~\ref{fig:Uncerts}. The predicted profile in this case is much more peaked, as can be seen from the green line in Fig.~\ref{fig:Uncerts}.
 To show a different case, we perform a similar test using the Nakanishi model, based on Ref.~\cite{Nakanishi_2003}.  
 Finally, as explained in the text, we also tested the convolution with a distribution that results from an even combination of the Ferriere and Nakanishi models, as a blue line (``Combined dist'' in the legend). 
 This, again, will simply decrease the fraction of DM contributing to the Galactic diffuse $\gamma$-ray emission in the regions where available measurements exist (the $|b|<15^{\circ}$, $|l|<30^{\circ}$ region, for instance). We remark here on the need for measurements of the diffuse emission at smaller regions around the GC to probe these kinds of profiles.
The last major ingredient that can affect the spatial distribution of the line here is how the different gas media are distributed and their temperature, which would change the factor $\sigma(E_\textrm{th})$ in Eq.~\eqref{eq:511line}. However, a detailed modeling of the different media and their temperatures is well beyond the scope of this work.

We estimate the average distance traveled by positrons in a neutral gas medium (considering that He constitutes around $10\%$ of the gas) as $\langle l \rangle = \sqrt{2 D(E) \tau_\textrm{loss}(E, n_\textrm{gas})}$, where D is the diffusion coefficient that we employ in this work and $\tau_\textrm{loss}$ is the loss-time given by the ionization energy losses. This is shown as blue lines in Fig.~\ref{fig:MeanFreePath}, where the solid lines represent the results for a $1$~cm$^{-3}$ medium and the dashed lines the same but for a $10$~cm$^{-3}$. We provide a rough estimate of the effect of reacceleration (green lines in Fig.~\ref{fig:MeanFreePath}) simply by evaluating the energy at which the positron spectra peak when including reacceleration (with our best-fit $V_A\sim14$~km/s) and assigning the average distance travelled by positrons of this energy. 
As one can see in Ref.~\cite{DelaTorreLuque:2023olp}, positrons injected by a $1-10$~MeV DM particle get boosted and their spectrum peaks at an average energy of a few tens of MeV, while for DM masses above $100$~MeV its effect becomes very minor.
Since this is an averaged distance, we remark that $50\%$ of the injected positrons actually will travel longer distances before thermalizing.
As a reference, a medium with density $n_\textrm{gas} = 100$~cm$^{-3}$ will travel an average distance $10$ times lower than in the $1$~cm$^{-3}$ case.

\begin{figure*}[b]
\includegraphics[width=0.49\linewidth]{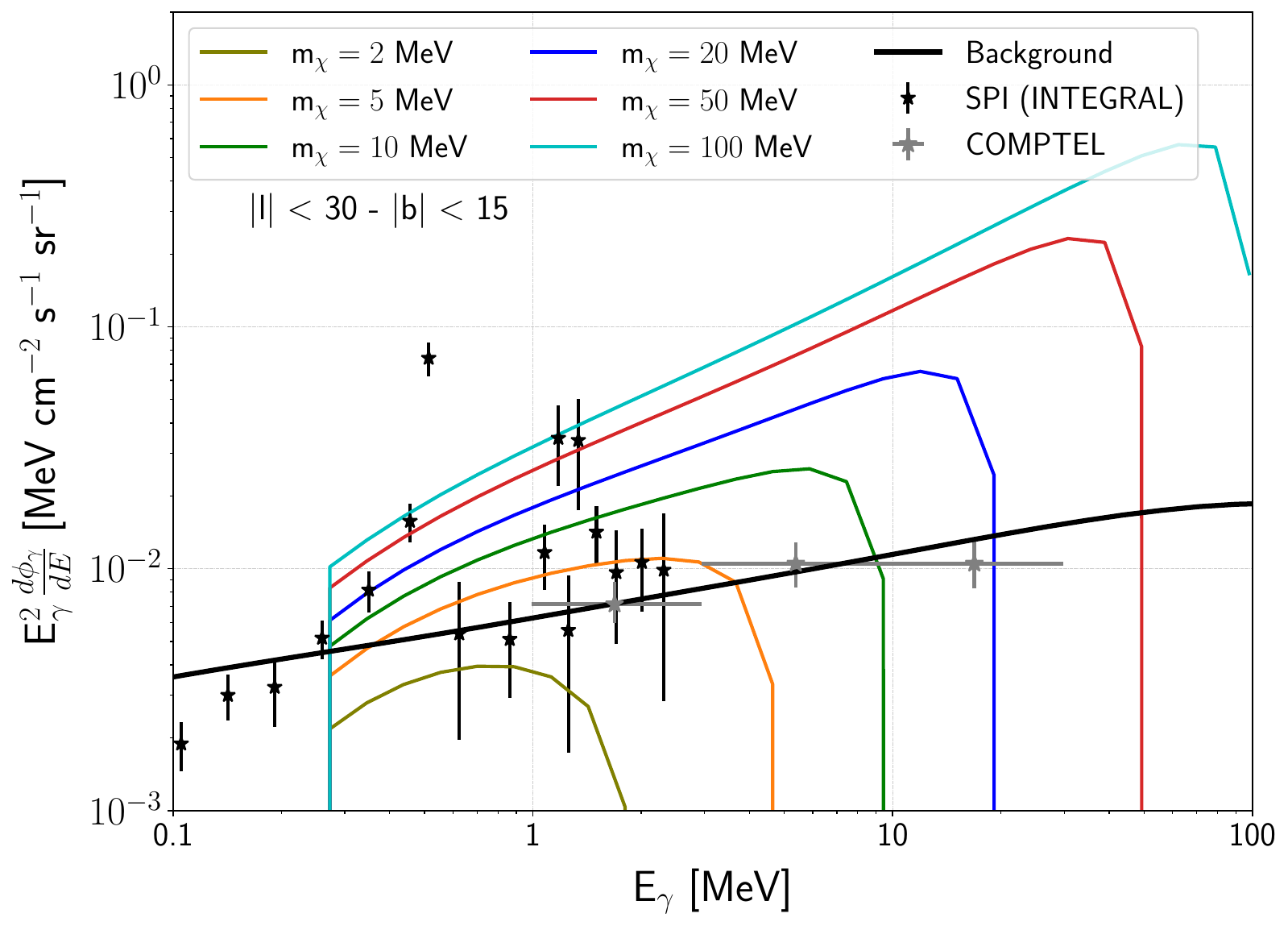}
\includegraphics[width=0.496\linewidth, height=0.248\textheight]{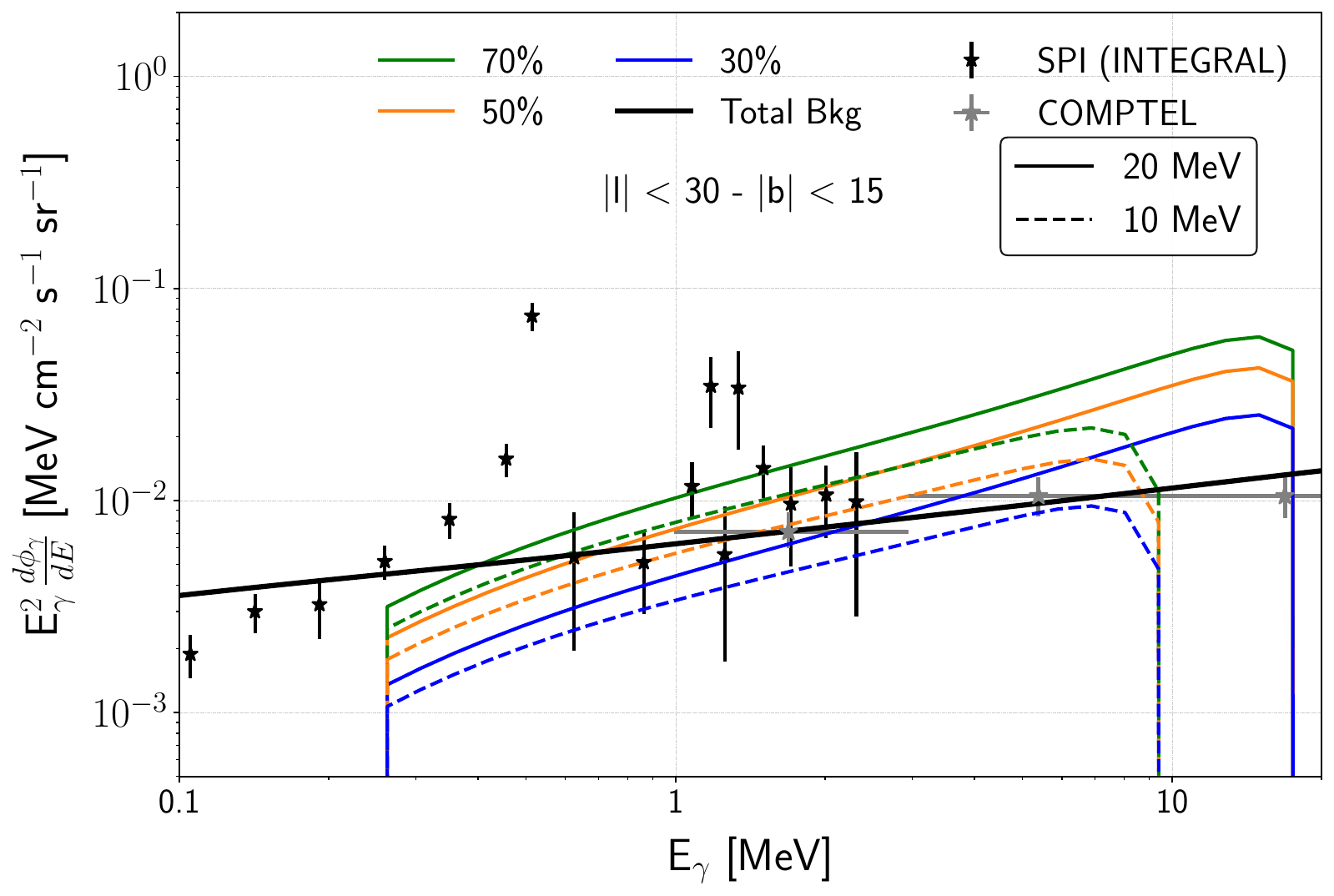}
\caption{Left panel: Predicted in-flight annihilation fluxes from DM particles with masses of $1$ to $100$~MeV, in the $|b|<15^{\circ}$ - $|l|<30^{\circ}$ region, compared to SPI measurements of the diffuse $\gamma$-ray emission. These predictions are normalized to the flux of the $511$~keV line in this region. Right panel: Expected flux from a source of monoenergetic positrons (positrons injected at $20$~MeV are shown as solid lines and those injected at $10$~MeV are shown as dashed lines) contributing to the total $511$~keV flux in the $|b|<15^{\circ}$ - $|l|<30^{\circ}$ region with the fractions shown in the legend. Both panels show the expected IC background contribution from CR electrons as a solid black line.}
\label{fig:IA_Em}
\end{figure*}

\section{In-flight annihilation emission from sub-GeV DM}
\label{sec:Scaling}
In this appendix, we show the predicted in-flight annihilation fluxes related to sub-GeV DM of different masses, normalized to produce the full $511$~keV emission in the $|b|<15^{\circ}$, $|l|<30^{\circ}$ region, where SPI has measured the continuum $\gamma$-ray flux up to energies around $3$~MeV. This serves as a cross-check with the results obtained from Ref.~\cite{Beacom:2005qv}. 
In the left panel of Fig.~\ref{fig:IA_Em}, we show the in-flight annihilation emission for DM particles with mass from $1$~MeV to $100$~MeV. As already shown in Fig.~\ref{fig:IA_Rat}, a larger DM mass implies a larger emission and up to higher energies. Without even accounting for any background emission, we observe that masses above $10$ GeV are ruled out by SPI data.  In the right panel of this figure, we also show the background IC emission predicted from Refs.~\cite{delaTorreLuque:2022vhm, DelaTorreLuque:2023zyd}. We note that a $5$~MeV DM particle already exceeds the background emission in this region. The sum of the background plus the in-flight contribution from a $3$~MeV DM particle exceeds the most constraining measurement (the data slightly below $1$~MeV) by $2\sigma$, in very good agreement with the findings of Ref.~\cite{Beacom:2005qv}

In general, one must take into account that only a fraction of the $511$~keV emission comes from high energy positrons, which are those generating large in-flight annihilation emission. In fact, Ref.~\cite{KeithHooper_511Lat} obtained strong constraints on PBHs from the fraction of the continuum emission at $511$~keV, meaning that this fraction is significant. In addition, one expects that the disk component must significantly contribute to the line, and other components are probably present too, as e.g. one is able to explain the asymmetry in the measurements of the longitude profile of the line. We show that the constraints on the mass of DM from in-flight annihilation must be significantly reduced when the DM-induced signal is only a fraction of the total $511$~keV flux. In the right panel of Fig.~\ref{fig:IA_Em} we show the in-flight annihilation emission for a $10$ (dashed lines) and $20$~MeV (solid lines) mono-energetic emitter contributing in different fractions to the total $511$~keV emission, from $30\%$ to $70\%$. These predictions are compared to SPI and COMPTEL~\cite{COMPTEL1994} data of the diffuse $\gamma$-ray emission. As we see, a source of $20$~MeV positrons could be compatible with the data if this source contributes only around $30\%$ of the total $511$~keV emission. Obviously, for the DM spike hypothesis that we propose here, the fraction of DM-induced emission over other contributions should be higher as we move to the GC. 

\begin{figure*}[t!]
\centering
\includegraphics[width=0.51\linewidth]{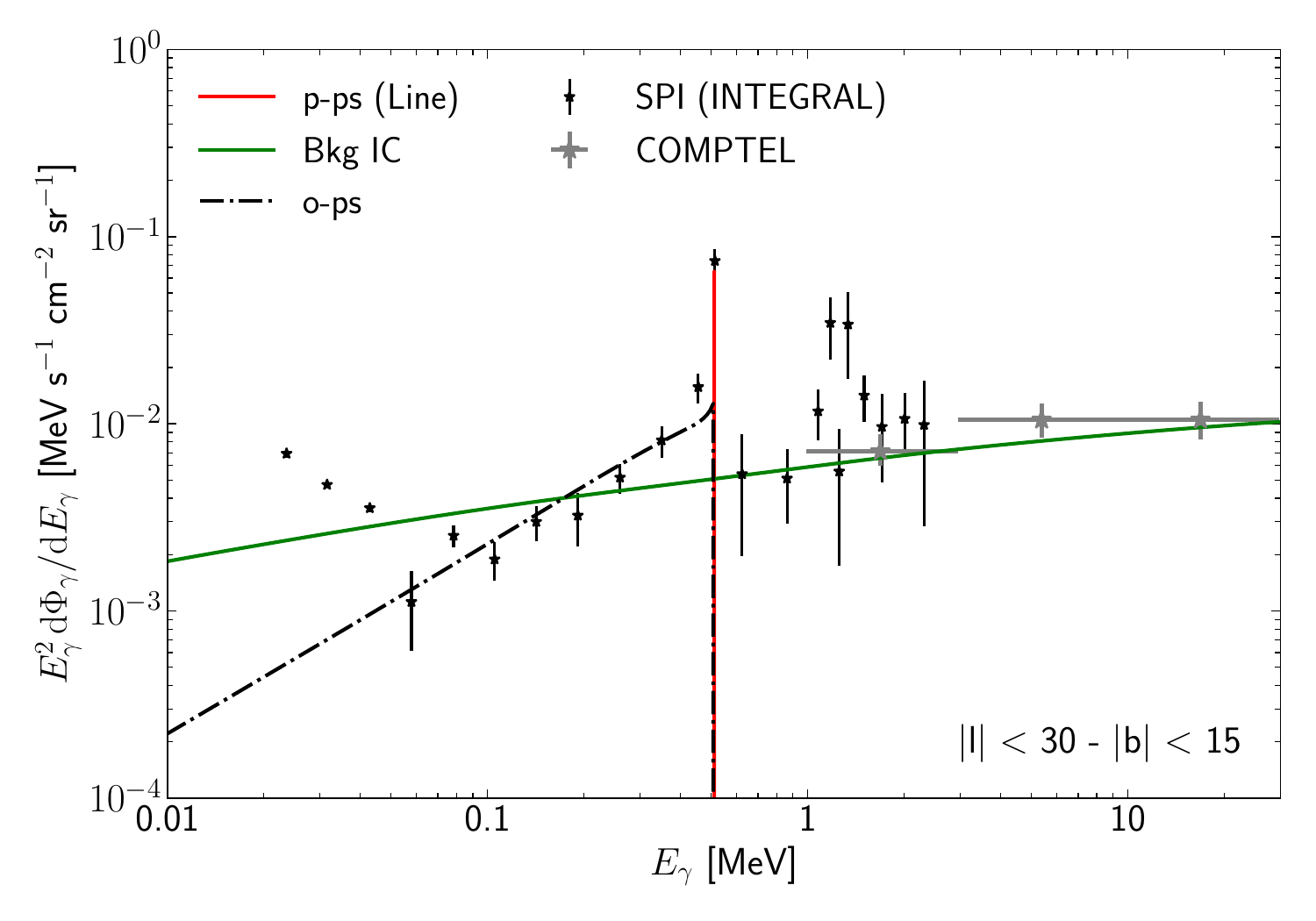}
\caption{Comparison of the expected background emissions independent of the DM mass with SPI and COMPTEL data in the $|b|<15^{\circ}$ - $|l|<30^{\circ}$ region. In particular, we show the IC background emission, the $511$~keV emission normalized at the measured value and the o-ps emission associated with the observed  $511$~keV emission.}
\label{fig:ops}
\end{figure*}

For completeness, we show in Fig.~\ref{fig:ops} the components of the continuum emission that are independent of the DM mass (or the energy scale of the injected positrons). In particular, we show the background emission, the $511$~keV line normalized to the observed value and the o-ps emission associated to this $511$~keV line emission. As we see, while the background emission matches the data above $511$~keV well, the o-ps emission already saturates the data below $511$~keV. This means that either the background IC must quickly drop exactly at $511$ keV, or it has a very hard (i.e. very steep) spectrum that would exceed COMPTEL data at higher energies. Therefore, this suggests that the emission measured from SPI below $511$~keV could be biased by the high systematic uncertainties in the measurement or by their template fitting analysis.

\begin{figure*}[t]
\includegraphics[width=0.496\linewidth, height=0.26\textheight]{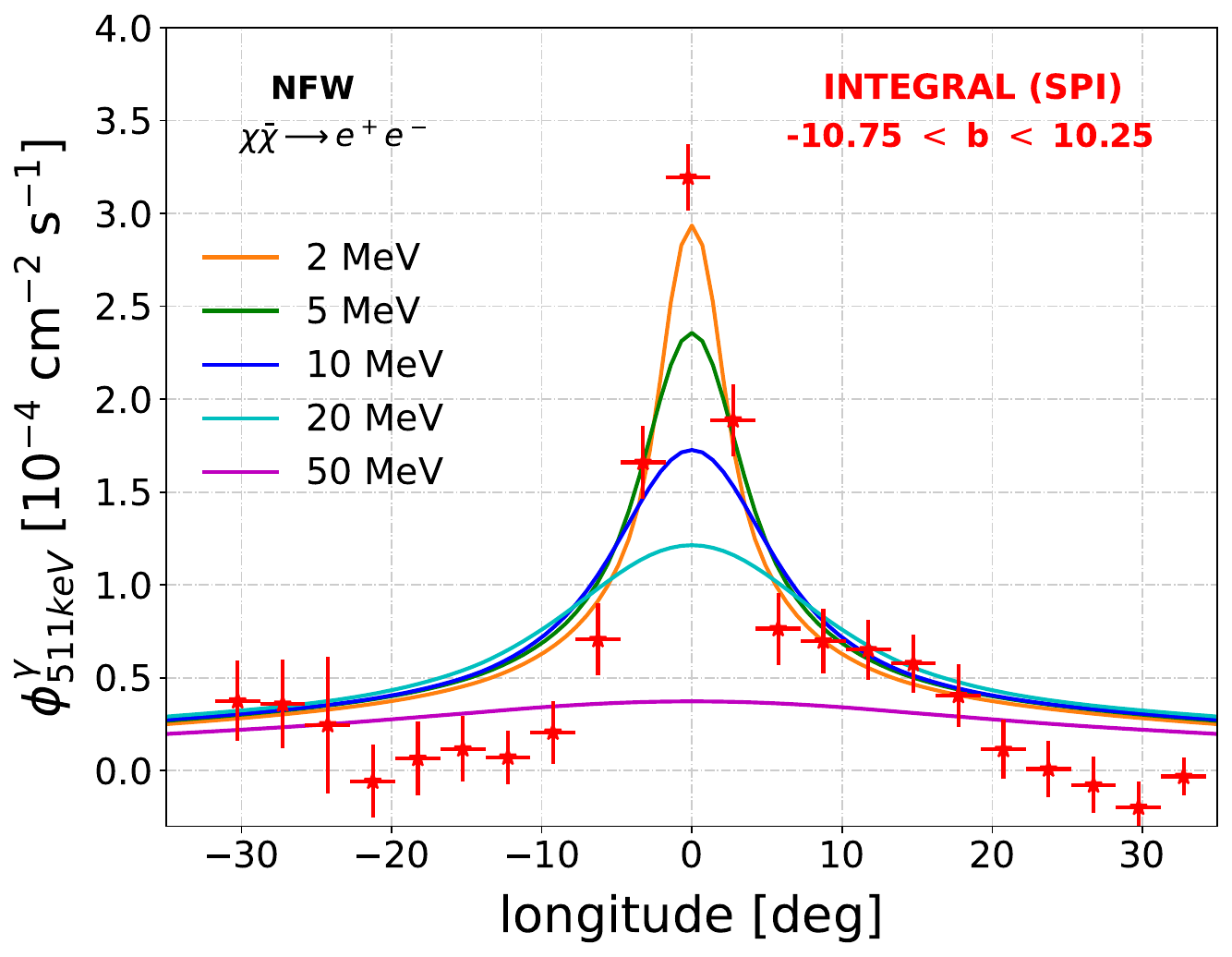}
\includegraphics[width=0.496\linewidth, height=0.26\textheight]{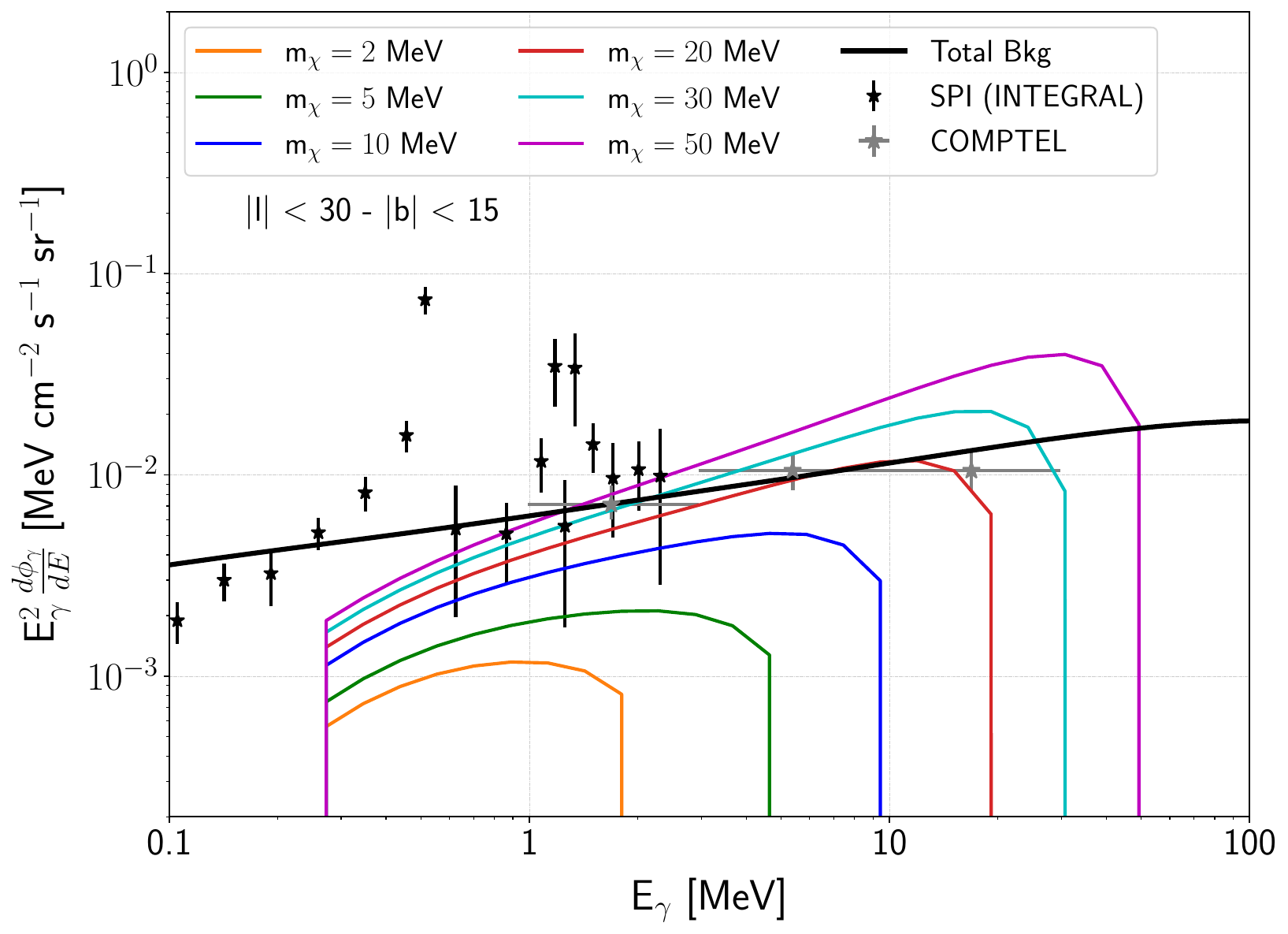}
\caption{Same as in Fig.~\ref{fig:Continuum_GS_SH} but for the NFW profile.}
\label{fig:NFWContinuum}
\end{figure*}

Finally, we also show the equivalent of Fig.~\ref{fig:Continuum_GS_SH} but for the NFW DM distribution in Fig.\ref{fig:NFWContinuum}. As we see, while the in-flight annihilation emission is similar to the other spike distributions, since it depends mostly on the mass of the DM particle, the predicted spatial morphology of the $511$ keV line distribution is unable to provide a good reproduction of SPI data. The corresponding $\langle \sigma v \rangle$ values are $4.1 \times 10^{-31}$~cm$^3$/s, $5.5 \times 10^{-30}$~cm$^3$/s, $2.6 \times 10^{-29}$~cm$^3$/s, $1.1 \times 10^{-28}$~cm$^3$/s and $8.4 \times 10^{-28}$~cm$^3$/s for DM masses of $2$~MeV, $5$~MeV, $10$~MeV, $20$~MeV and $50$~MeV, respectively. 


\begin{figure*}[b!]
\includegraphics[width=0.5\linewidth]{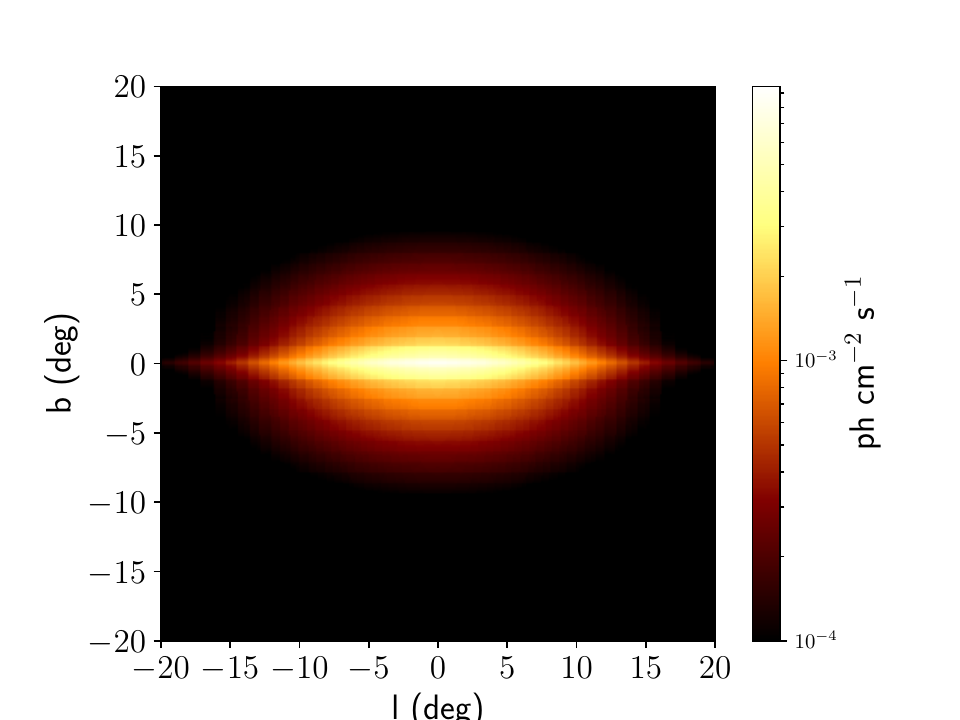}
\includegraphics[width=0.5\linewidth]{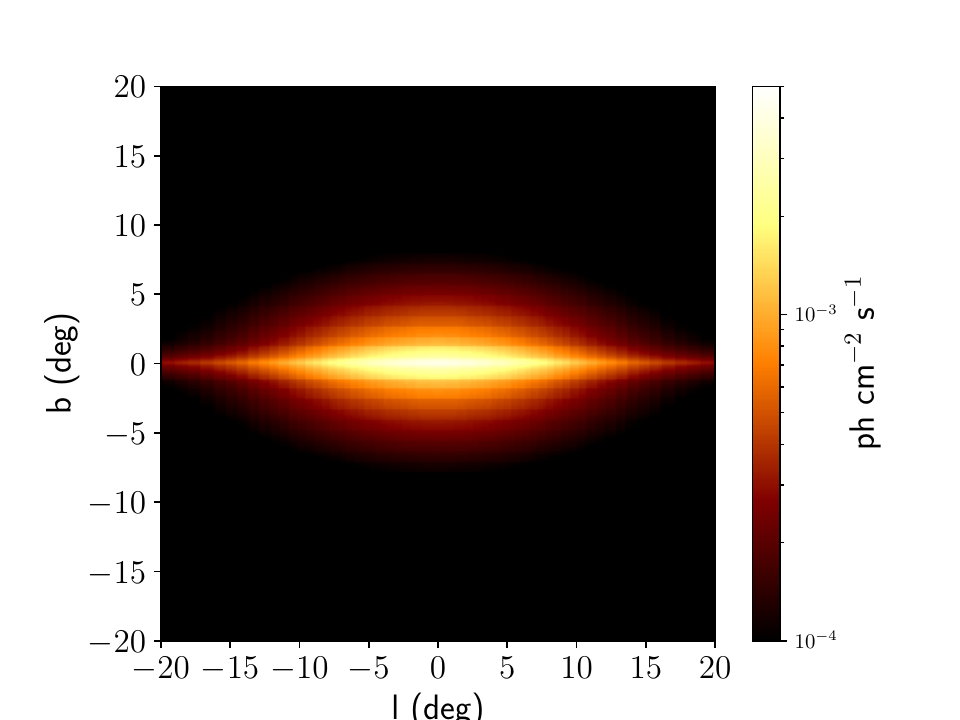}
\caption{Maps of the in-flight annihilation emission around the Galactic Center, for a 20 MeV DM mass and cross sections of $10^{-32}$~cm$^3$/s and $10^{-30}$~cm$^3$/s, for the Gondolo-Silk profile (left panel) and *Min profile (right panel), respectively.}
\label{fig:Maps}
\end{figure*}

\section{Other regions of interest}
\label{sec:ROIs}
This appendix is intended to compare the predicted in-flight positron annihilation emission with other publicly available observations of the diffuse MeV Galactic emission, especially by COMPTEL. 

First, we show the maps of the expected in-flight annihilation emission in the inner $20^{\circ}$, in Fig.~\ref{fig:Maps}, for a 20 MeV DM mass and cross sections of $10^{-32}$~cm$^3$/s and $10^{-30}$~cm$^3$/s, for the Gondolo-Silk profile (left panel) and *Min profile (right panel), respectively. As we can see, the effect of the exponential suppression of the electron density with latitude is evident in both cases.

\begin{figure*}[b!]
\includegraphics[width=0.49\linewidth]{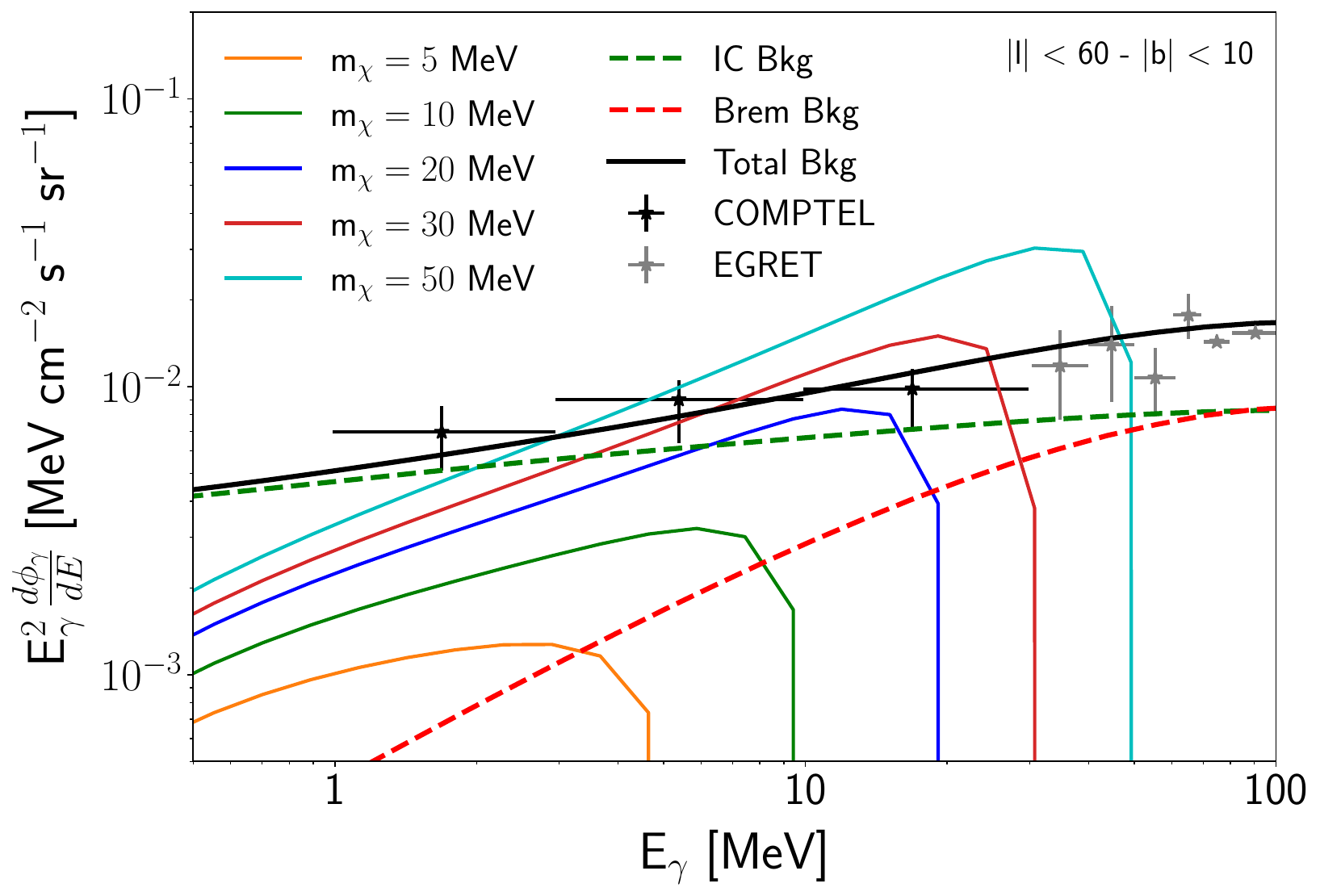}
\includegraphics[width=0.496\linewidth]{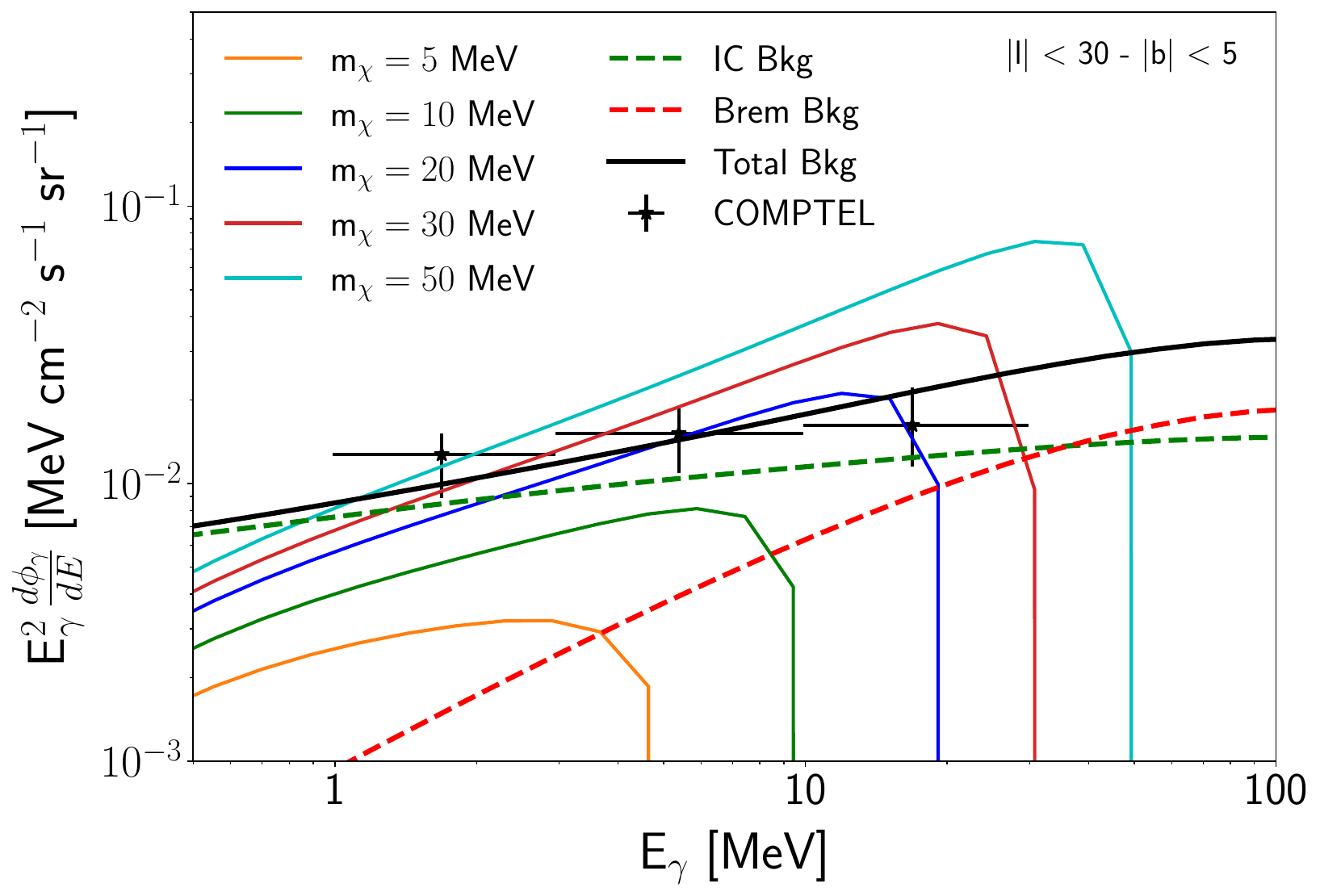}

\includegraphics[width=0.49\linewidth]{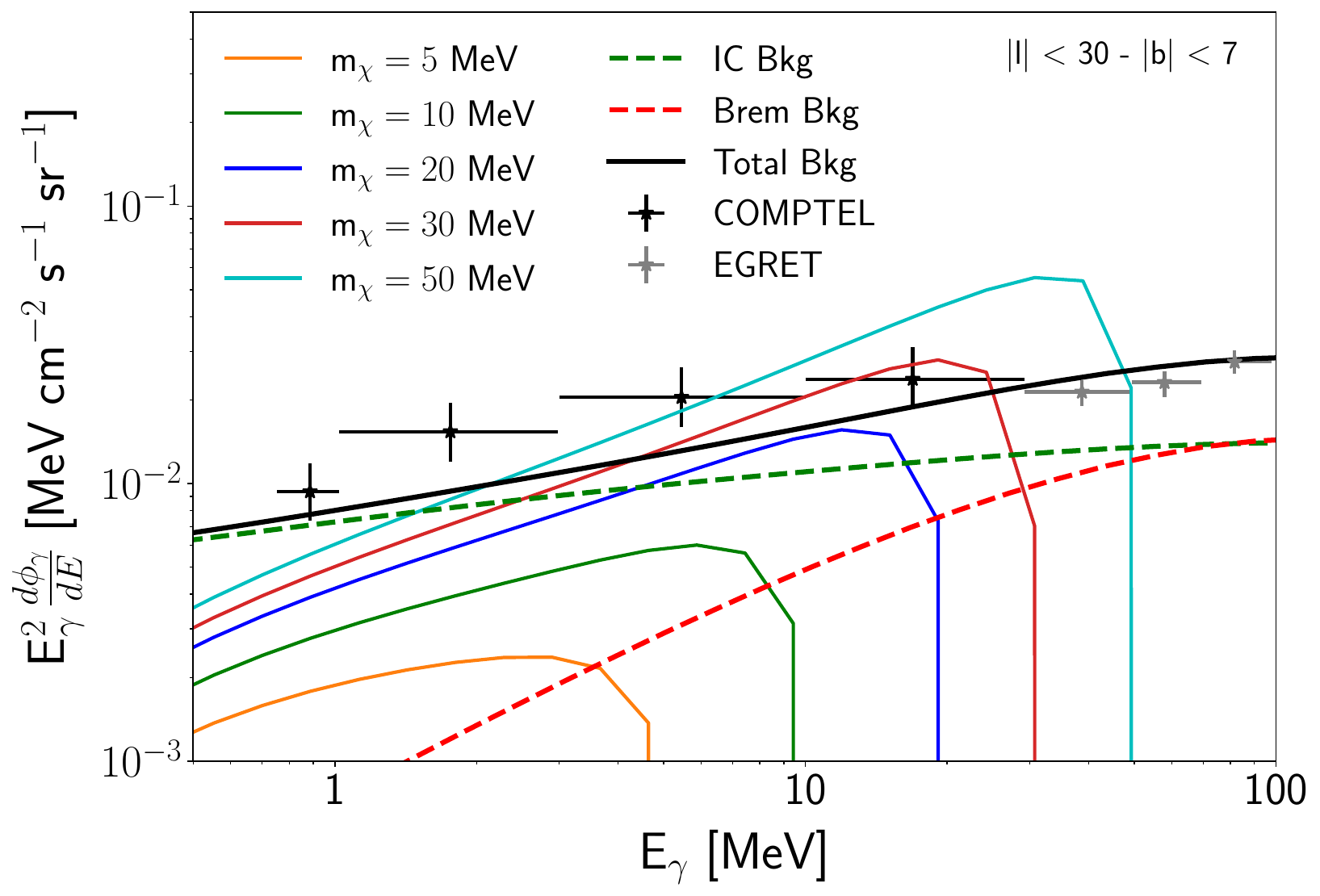}
\includegraphics[width=0.496\linewidth]{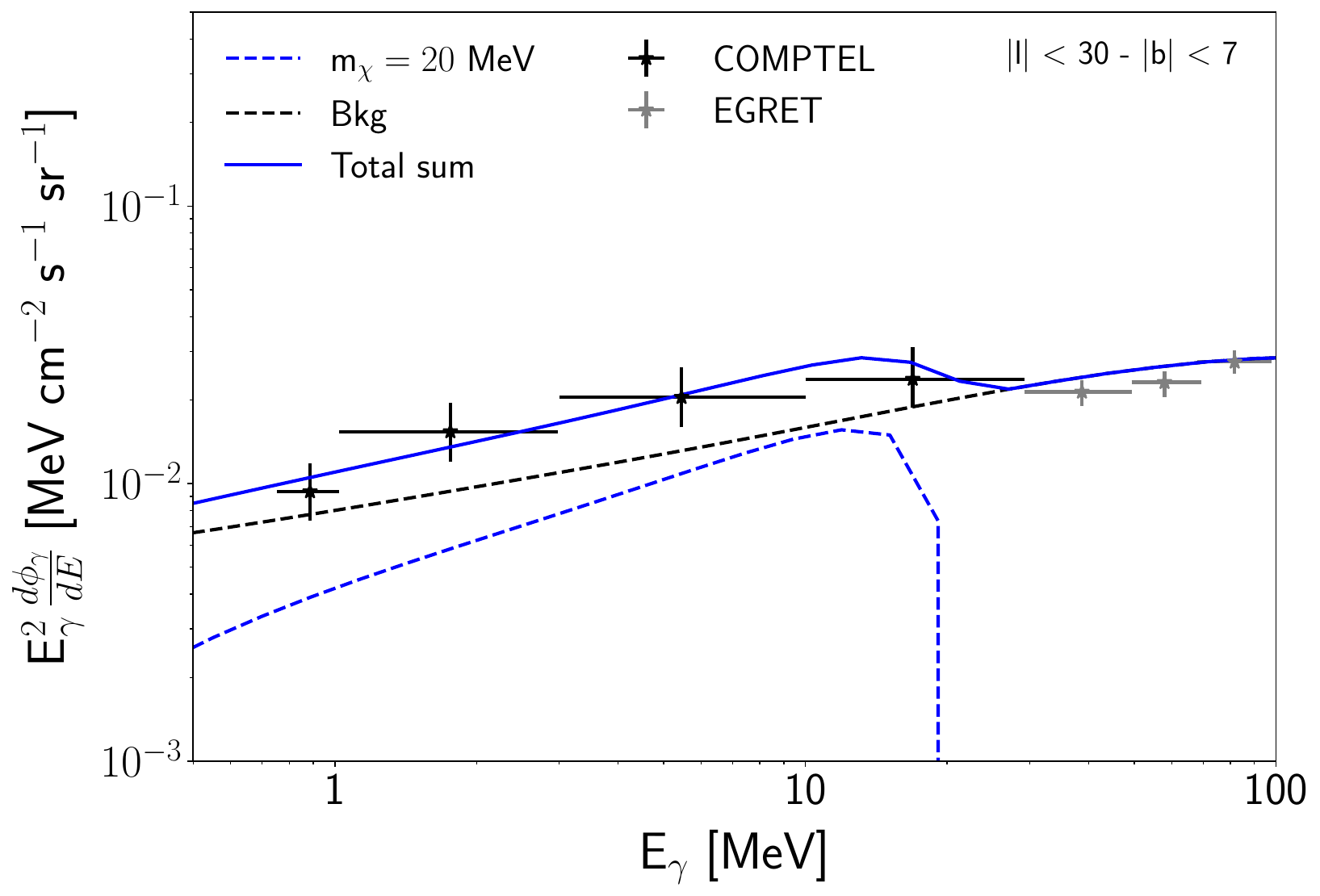}
\caption{Predicted in-flight annihilation flux for various DM masses and assuming a Gondolo-Silk profile, with cross sections obtained from the fit of SPI longitude profile of the $511$~keV line. The background components (bremsstrahlung and IC) are shown as dashed lines and the total background emission is shown as a solid black line. These are compared with COMPTEL and EGRET measurements of the Galactic diffuse emission in the $|b|<10^{\circ}$ - $|l|<60^{\circ}$ (top-left panel) region, $|b|<5^{\circ}$ - $|l|<30^{\circ}$ (top-right panel) region and the $|b|<7^{\circ}$ - $|l|<30^{\circ}$ (bottom panels) region. For clarity, we depict, in the bottom-right panel, a comparison of the total emission (background + in-flight annihilation) with the data for a $20$~MeV DM particle, that seems to show a hint indicating the need of an in-flight annihilation signal to reproduce the observations.}
\label{fig:ROIs}
\end{figure*}

Then, we focus on the comparisons of the expected in-flight positron annihilation emission signals for other ROIs. In these comparisons, the in-flight positron annihilation fluxes are calculated in a similar fashion to the right panels of Fig.~\ref{fig:Continuum_GS_SH} (i.e. taking the cross sections that provide the best-fit of the $511$~keV line emission to the longitude profile of SPI data). These comparisons are illustrated in Fig.~\ref{fig:ROIs}, that show the predicted emissions for a Gondolo-Silk profile with COMPTEL and EGRET measurements of the Galactic diffuse emission in the $|b|<10^{\circ}$ - $|l|<60^{\circ}$ (top-left panel), $|b|<5^{\circ}$ - $|l|<30^{\circ}$ (top-right panel) and the $|b|<7^{\circ}$ - $|l|<30^{\circ}$ (bottom panels) region. In addition, the background components (bremsstrahlung and IC) are shown as dashed lines and the total background emission is shown as a solid black line.

From this figure, we observe that the in-flight annihilation signals are compatible with the data for masses even slightly larger than $20$~MeV, this is even when summing with the background emission, which coincides with what we see from Fig.~\ref{fig:Continuum_GS_SH}.
In addition, as in the case of the $|b|<15^{\circ}$ - $|l|<30^{\circ}$ region shown in the main text, COMPTEL data seems to indicate a background emission going roughly as $\sim E^{-1.9}-E^{-1.8}$, which is compatible with what our background model predicts. We remind the reader that this background emission comes from a combined fit of the local electron CR spectrum at $\sim$GeV energies with the local emissivity data from Fermi-LAT~\cite{delaTorreLuque:2022vhm, DelaTorreLuque:2023zyd}. We also point out that the diffuse background emission measurements from Refs.~\cite{Siegert_2022, Berteaud_2022} using SPI observations are in good agreement  with this background model, as shown in Fig.5 of Ref.~\cite{luque2022optimizedflukacrosssections}.

Remarkably, we observe that in all the regions where the background emission model is in agreement with the data, there is a region where the data shows an excess over the predicted background. This can be well described with an extra  bump-like component. In fact, this extra component seems to match very well with a in-flight annihilation signal of a DM particle with a mass of about $20$~MeV. This may be the first indication of a hint favoring an in-flight annihilation signal, and we plan to study it in detail in a future work.

\begin{figure*}[t!]
\includegraphics[width=0.49\linewidth]{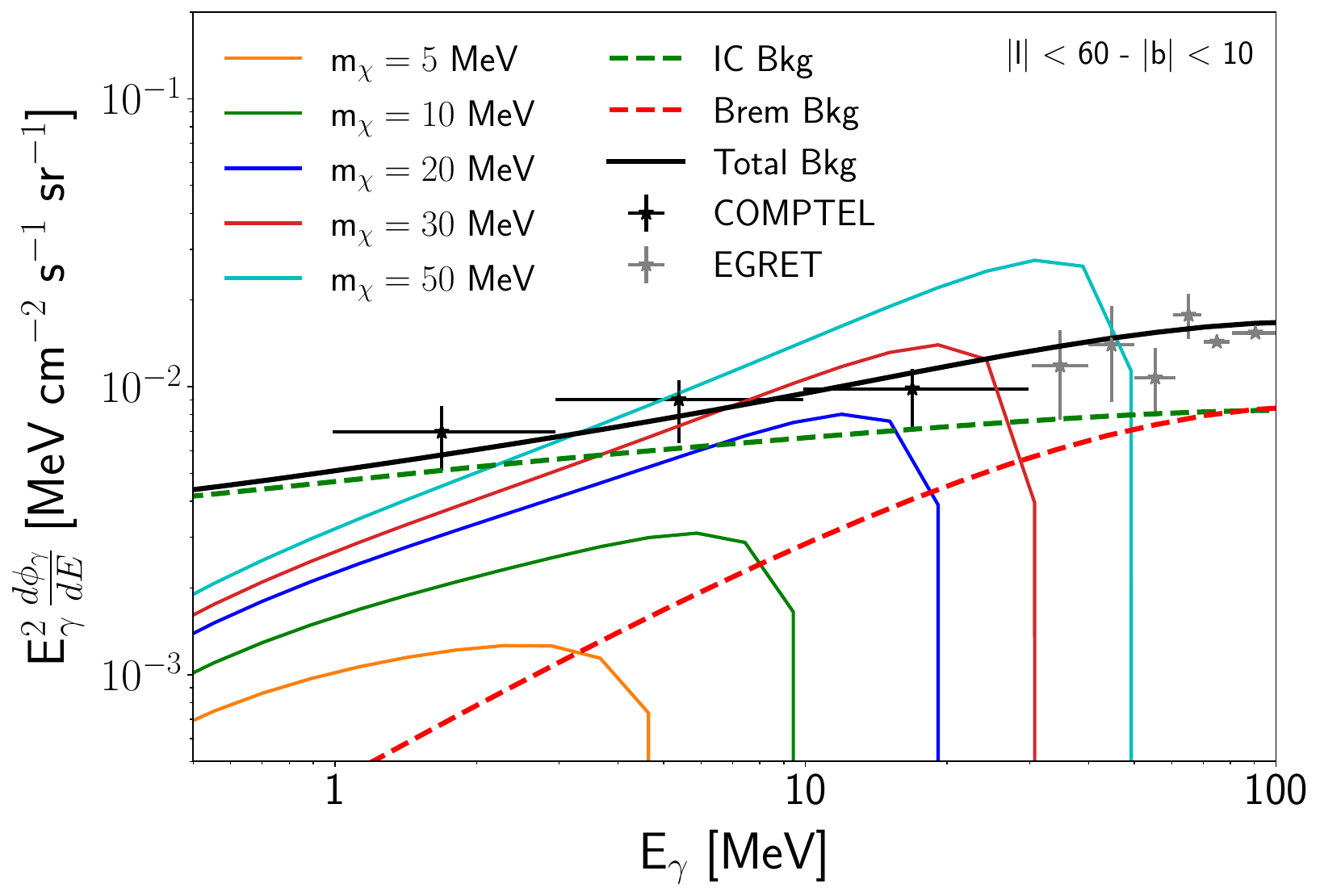}
\includegraphics[width=0.496\linewidth]{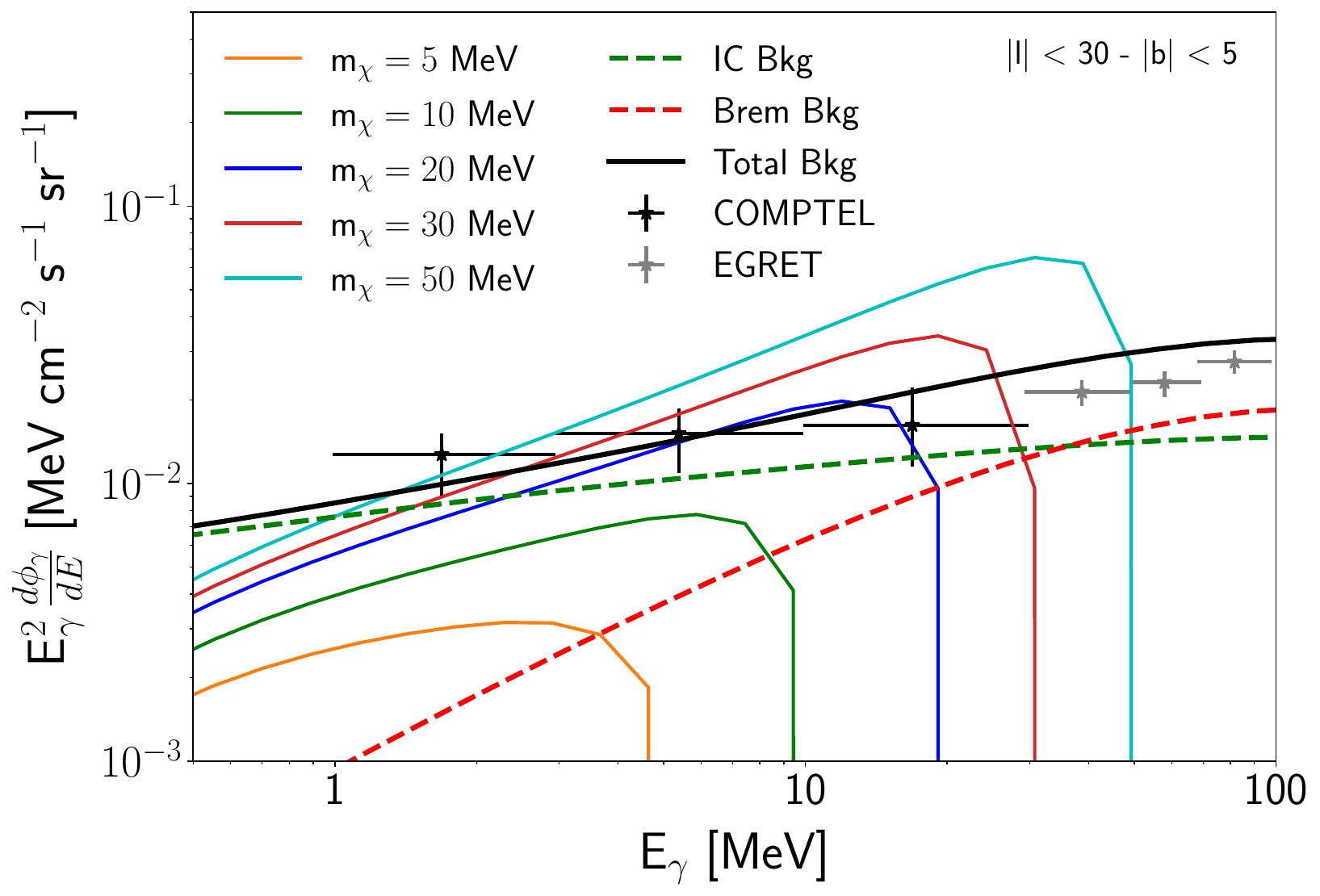}

\includegraphics[width=0.49\linewidth]{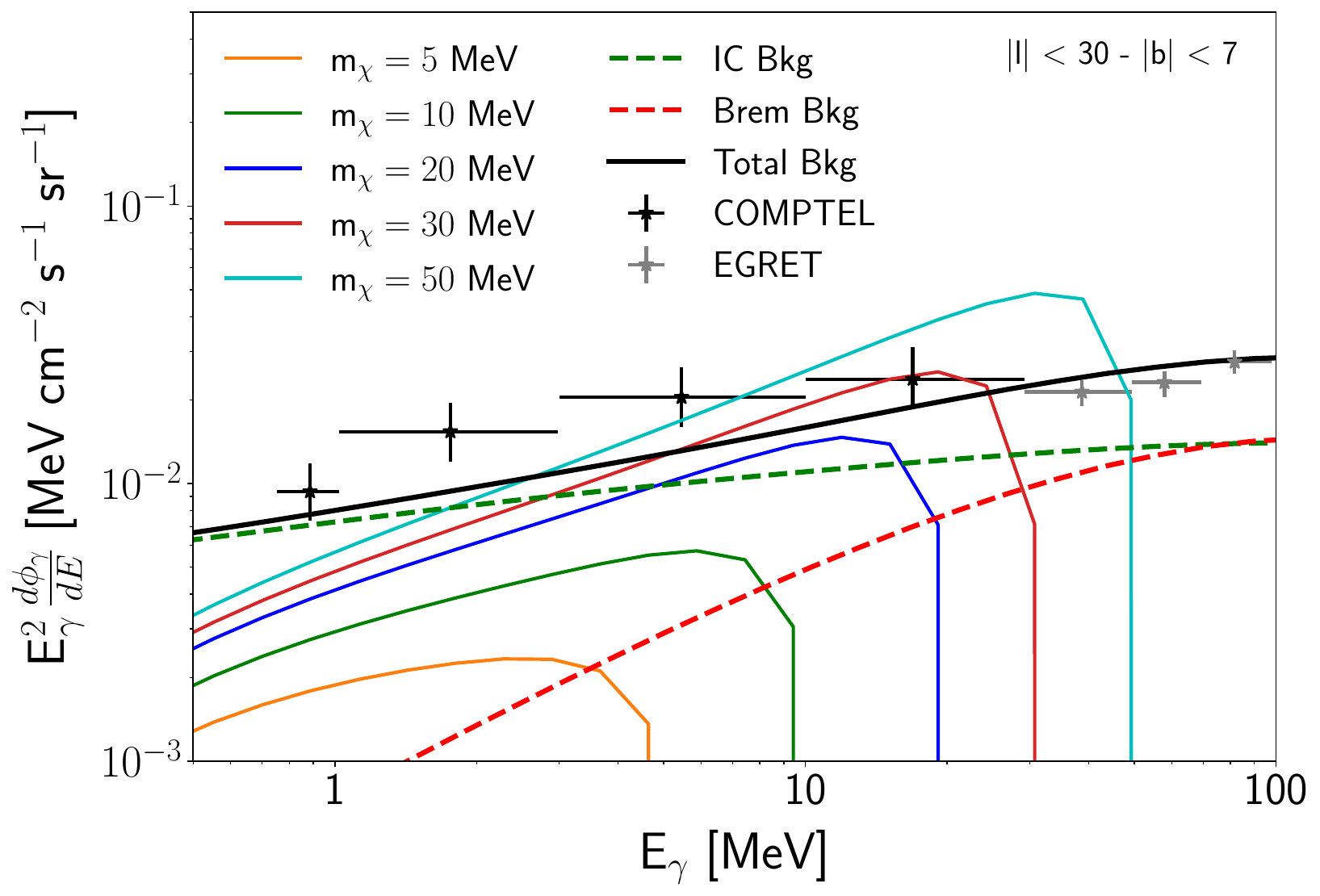}
\includegraphics[width=0.496\linewidth]{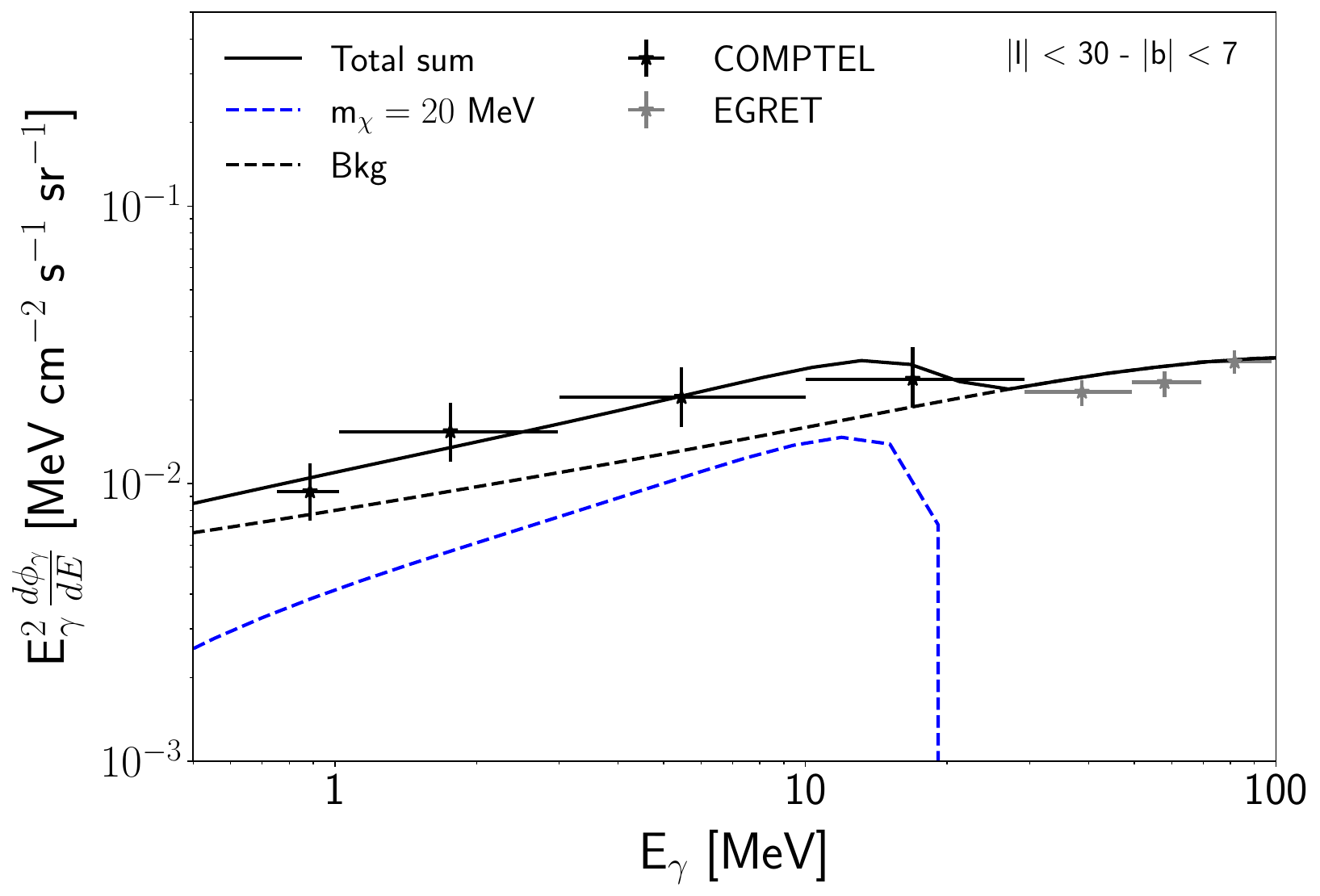}
\caption{Analogous to Fig.~\ref{fig:ROIs} but for the *MIN spike model. In the bottom-right panel, we show a comparison of the total emission with measurements considering a $20$~MeV DM particle.}
\label{fig:ROIs_SH}
\end{figure*}

\begin{figure*}[t!]
\includegraphics[width=0.49\linewidth]{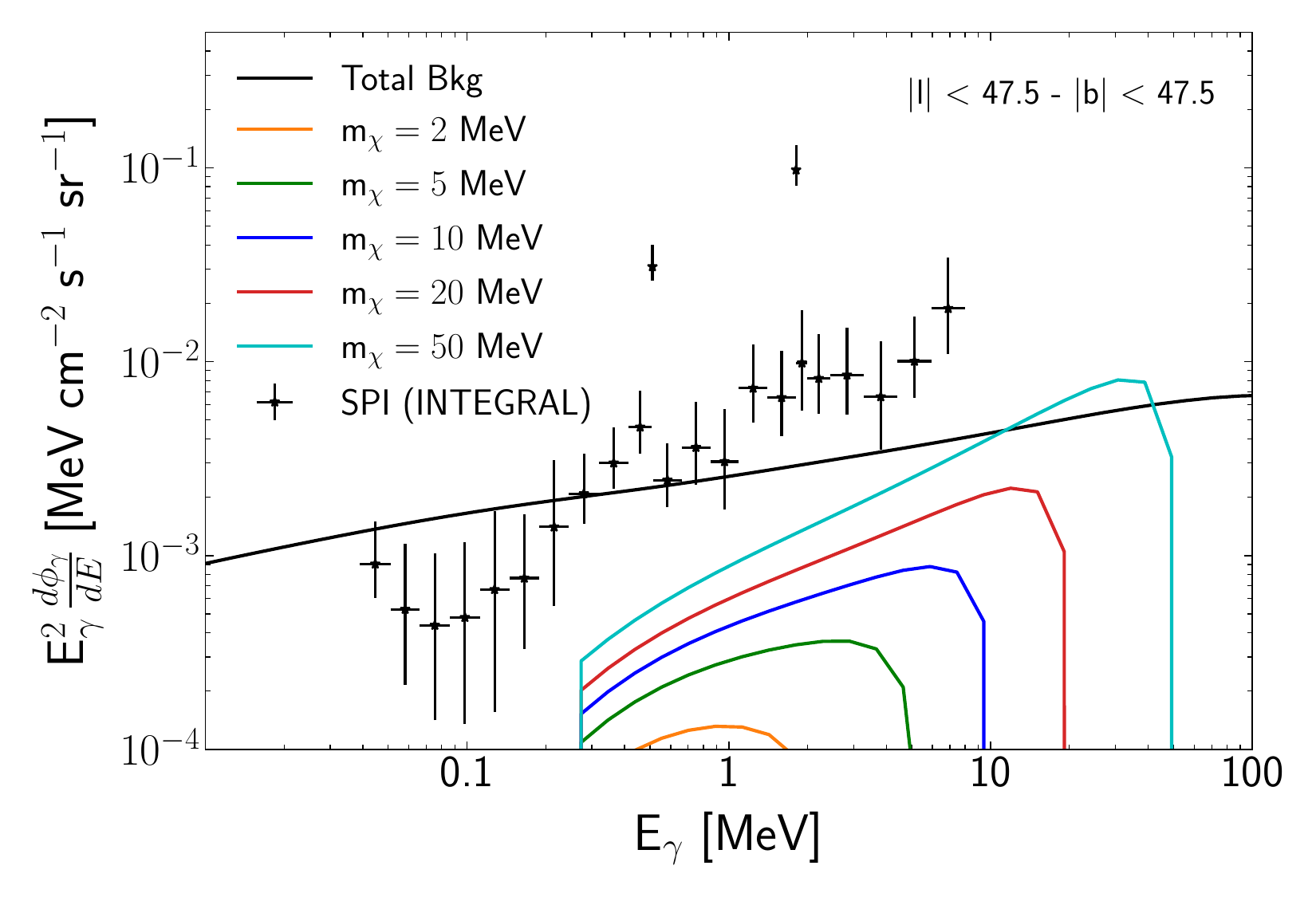}
\includegraphics[width=0.496\linewidth]{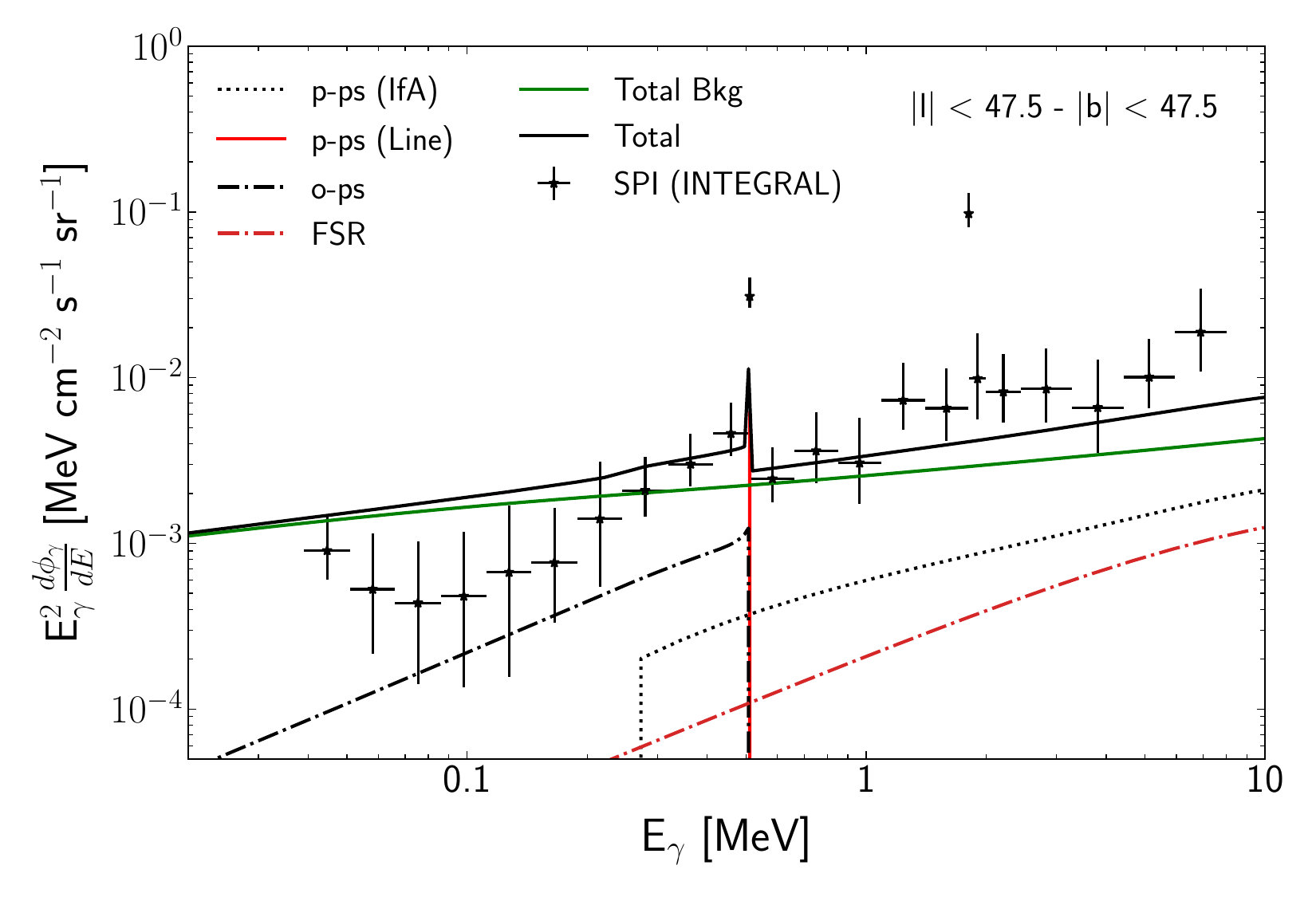}

\includegraphics[width=0.49\linewidth]{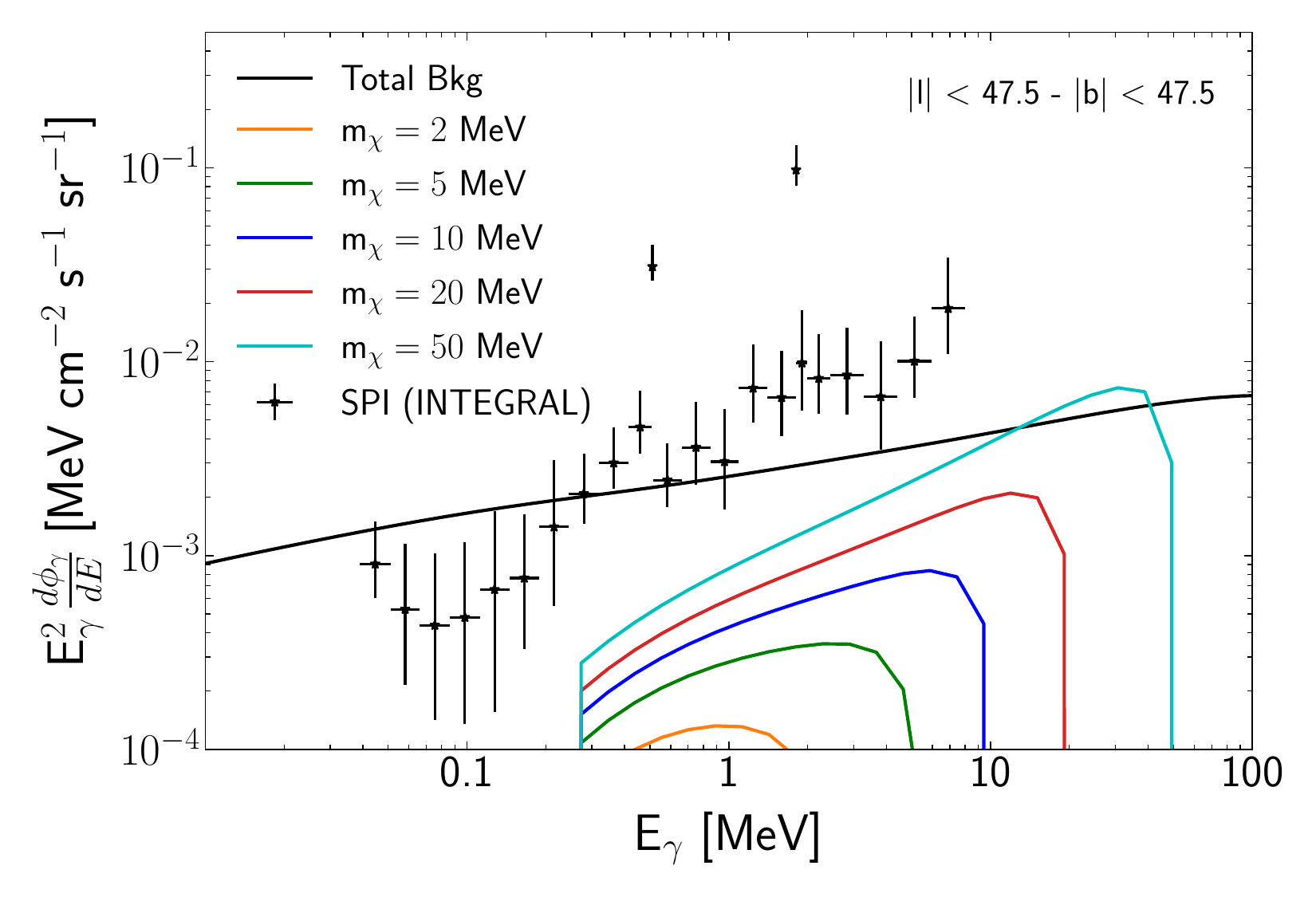}
\includegraphics[width=0.496\linewidth]{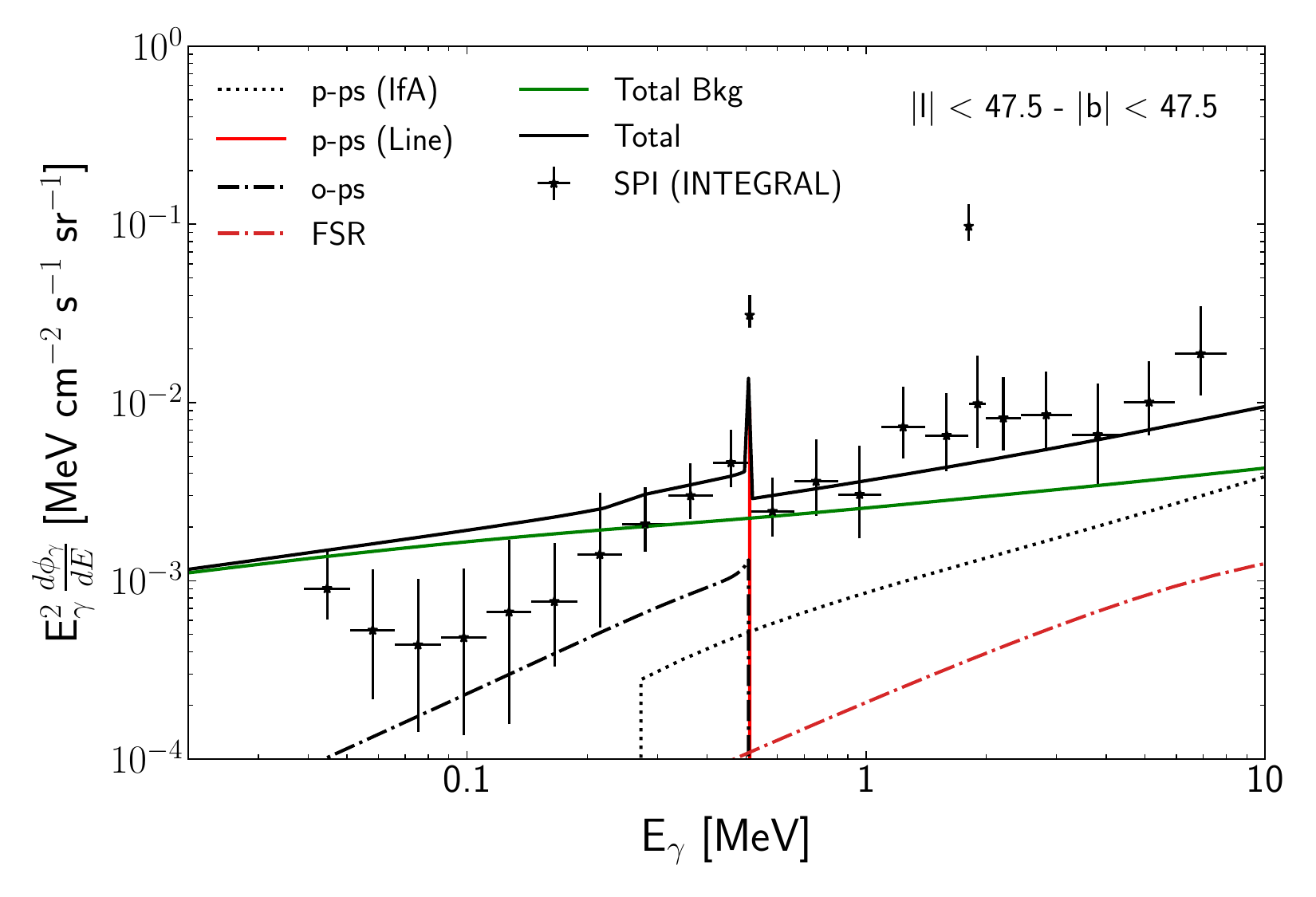}
\caption{IA signals for different DM masses compared to SPI data in the $47.5^{\circ}$ inner Galactic region. Top panels represent the predicted emission from the Gondolo-Silk profile while the bottom panels refer to the *Min profile. Analogous to Fig.~\ref{fig:Continuum}, in the righ panels we show all the expected contributions to the continuum gamma-ray flux from a $20$~MeV DM particle, summing also with our model for the background IC emission.}
\label{fig:IA47.5}
\end{figure*}

Finally, we also show in Fig.~\ref{fig:IA47.5} similar comparisons but for the SPI measurements in the $47.5^{\circ}$ around the GC. As it is expected, the contribution to the 511 keV line emission and the continuum emission is much lower than in more central regions of interest.

\end{document}